\documentclass[runningheads]{svmult}

\usepackage{makeidx}   
\usepackage{graphicx}  
\usepackage{subeqnar}  
\usepackage{multicol}  
\usepackage{physprbb}  
\makeindex             

\def\ni{\noindent }
\def\etal{{\sl et al.}}

\def\y3{{v}}

\usepackage{amsfonts}   
\usepackage{epsfig}  
\begin{document}
\title*{Seismic diagnostics for rotating massive  main sequence  stars}
%
%
%
%
\titlerunning{ }
%

\author{Mariejo Goupil}
\authorrunning{Mariejo Goupil}

\institute{Observatoire de Paris,  \\ 5 place Jules Janssen, 92190 Meudon principal cecex, France \\
E-mail: \texttt{mariejo.goupil@obspm.fr}}


\maketitle              

\begin{abstract}
\noindent Effects of stellar rotation on adiabatic oscillation frequencies
 of $\beta$ Cephei star 
  are discussed. 
Methods to evaluate them are briefly described and  
some of the main results  for four specific stars are presented. 
\end{abstract}


\section{Introduction} 

Main sequence (MS) massive stars are usually fast rotators and their fast rotation affects 
their internal structure  as well as their evolution.
The issue which is adressed here is what 
information can we obtain- about rotation - from the oscillations 
of these  massive, main sequence stars ?

The following seismic diagnostics for rotation using non axisymmetric
 modes will be discussed: 
 {\it 1) rotational splittings}  as  direct   probes of     the rotation profile.
 More precisely, we study the  effects of cubic order 
 in the rotation rate 
compared to effects of a  latitudinal dependence of the
 rotation on the splittings; 
 {\it 2) splitting asymmetries}  as a probe for centrifugal distorsion.
The case of  {\it 3) axisymmetric modes}  as   indirect probes   of rotation
throughout  effect of rotationally induced mixing on the structure will also be
considered.

Results discussed here are obtained with perturbation methods. 
For nonperturbative methods and
results, we refer the reader to Ligni\`eres {\sl et al.} (2006),  Reese {\sl et al.} (2009)
and references  therein.

The paper is organized as follows: in Sect.2, properties of 
pulsating B  stars are recalled with emphasis on the uncertainties of their physical
 description that can be
 addressed by seismic analyses. 
Sect. 3. recalls the  theoretical framework for seismic analyses of relevance
here. In Sect.4,
 seismic analyses of four  $\beta$ Cep  are presented. In Sect.5
 a theoretical  study compares the modifications of the rotation splittings
 due either to
 latitudinal dependence of the rotation rate, $\Omega$,
 or to cubic order ($O(\Omega^3$)) frequency corrections. 
 Some conclusions are given in Sect.6.

\section{Massive main sequence stars}\label{MMS} 

O-B stars are  characterized by a convective core  and an envelope
 which is essentially radiative apart a thin  outer  region related to 
  the  iron opacity  bump. 
 Important incertainties regarding the structure   and future evolution of these stars 
 are: \\
-the extent of chemical element mixing beyond the central instable
 layers as defined by the Schwarzschild criterium \\
-Transport of angular momentum because the rotation 
can  play  a significant  role in chemical element mixing

\begin{figure}[t]
\centerline{\epsfig{file=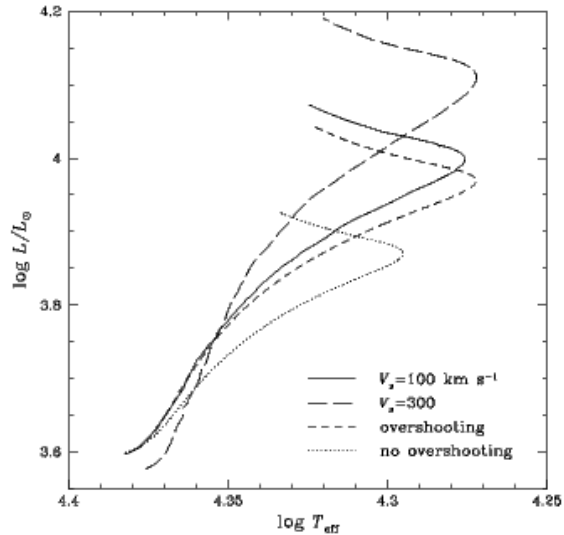,width=8cm}}
 \caption{Evolutionary  tracks for   $9 M_\odot$  models with neither rotation, nor
 overshoot included (dotted line), with overshoot included  but not rotation
 (short dashed line)
 and  with rotation included but   not overshoot (long dashed and solid lines)
{\it (from Talon {\sl et al.}, 1997)} }
\label{Suzanne1} 
\end{figure}

{\it Convective core overshoot:}
In 1D stellar models, the convective core is delimited by the  radius $r_{zc}$
according to the Schwarzschild criterium  $\nabla_{ad} = \nabla_{rad}$. However this corresponds to 
a vanishing buoyancy
force: the eddies are then strongly slowed down but still retain some velocity.
Hence due to  inertia, eddies move beyond 
the Schwarzschild radius till their velocity vanishes  that-is 
 over a distance $d_{ov}$ such that the effective convective core radius becomes  $r_{ov}=r_{zc}+d_{ov}$.
Despite theoretical  investigations (Zahn, 1991, Roxburgh, 1992),
 the overshooting distance computed in  1D 
  stellar evolutionary models  usually remains a rough prescription i.e. it is assumed that 
   $d_{ov} = \alpha_{ov} {\rm min}(r_{zc}, H_p)$   with  
  $H_p$  the local pressure scale height and $\alpha_{ov}$ is a  free parameter. Empirical
  determinations from observations  (Schaller {\sl et al.} 1992, Cordier {\sl et al.} 2002;
   Claret, 2007 and references therein) 
  yield a wide range for  $\alpha_{ov}$, namely [0-0.5] $H_p$.   
The adopted value for this free parameter has important  consequences for 
the  evolution of a model  with a given mass: 
with a higher luminosity, it is  older at given central hydrogen content ($X_c$) 
on the MS and reaches the end of the MS with a    larger mass core- total mass ratio.  
On a statistical point of view,  the value of  $\alpha_{ov}$ affects the thickness of the MS on a HR diagram 
as well as the isochrones.   Core overshoot has therefore an influence on stellar age
determination (Lebreton {\sl et al.} 1995, Lebreton, 2008).  

 \medskip

{\it Rotationally induced mixing in radiative regions:}
Departure from thermal equilibrium generated by the
oblateness of a rotating star  causes large scale motions, the meridional circulation.  As   
differential rotation also  induces turbulence,  competition of these two processes can result 
in (rotationally induced)  diffusion of chemical elements (Zahn, 1992 and subsequent works).
The evolution of a given chemical specie $j$ with concentration $c_j$
  is governed by a diffusion equation  
  (for a review, Talon, 2008; Decressin {\sl et al.} 2009):
\begin{equation} 
\rho {dc_j\over dt} = \rho \dot c_j + {1\over r^2} {\partial \over \partial r}
[r^2 \rho V_{ip}]  +   {1\over r^2} {\partial
\over \partial r} [r^2 \rho D_t{\partial c_j \over \partial r}] 
\label{cj}
\end{equation} 
where the first term represents nuclear transformation and
 the second term atomic diffusion with $V_{j,p}$ the diffusion velocity of
 particles j with respect to protons
and where the tuirbulent diffusion coefficient 
$D_t= D_{eff}+D_v $,  
$D_{eff}$ comes from the meridional circulation and $D_v$ from the turbulence.
As $D_{eff}$ depends on the vertical meridional velocity $U_r$, 
chemical and angular momentum evolutions must be solved together. Hence one also solves
an (diffusion-advection) evolution equation for the angular momentum :
\begin{equation} \rho {d r^2 \Omega \over dt} = {1\over 5 r^2} {\partial \over \partial r} [r^4 \rho \Omega U_r]  +  
 {1\over r^4} {\partial\over \partial r} [r^4 \rho \nu_v {\partial \Omega \over \partial r}]
 \end{equation}  
 where $\nu_v $ is the vertical turbulent viscosity related to rotational
 instabilities. 
The current  picture is that the vigor  of the  meridional circulation is controlled by the magnitude  
of the surface  losses of angular momentum. Hence for hot, high mass stars which lose mass but
much less angular momentum, one expects 
 no efficient angular  momentum internal transport. The 
 rotation profile then essentially results  from
 expansion and contraction within the star during its evolution:
  i.e. high ratio of core rotation over surface
 rotation. This is well reproduced by rotationally induced mixing of type I 
 (Talon {\sl et al.} 1997). 
 On the other hand, for cool stars with extended convective outer layers, 
 dynamo generates an  efficient 
 magnetic driven wind  which is  efficient to drive important angular momentum losses and 
 internal transport.   This mechanism however is not
 sufficient enough in the solar case to make the observed rigid rotation in the radiative solar 
 interior and one must calls for to  other mechanisms (waves, magnetic field)
  (see Talon, 2007; Rieutord, 2006 for reviews). 
  This shows that many open questions related to stellar  internal rotation and its gradients subsist.  
 An important issue then is    to locate  regions of uniform rotation  
 and   regions of  differential rotation  (depth and/or latitude dependence)
inside the star   ($\Omega_{core}/\Omega_{surf}$). Another 
problem which must be solved  is 
 to disentangle effects of overshooting and rotation on  mixed  
 central  regions  and extension of convective cores.  Indeed the rotationally induced chemical 
mixing  affects the  evolution  of the star, its internal structure and oscillation
frequencies as  does core overshoot   although in a different way
(Talon {\sl et al.}, 1997; Goupil \& Talon, 2002; Montalban  {\sl et al.}, 2008; 
Miglio {\sl et al.}, 2008; Thoul, 2009). Fig.\ref{Suzanne1}  illustrates 
  the respective effects of element mixing by core overshoot and 
rotation   on the evolution  of a 9 $M_\odot$ main sequence model in a HR diagram.  

Seismology of  O-B stars can bring some light about these processes. More specifically, 
 $\beta$ Cephei  stars are good candidates for this purpose 
 (Montalban  {\sl et al.}, 2008; 
Miglio {\sl et al.}, 2008; Goupil \& Talon, 2008; Miglio {\sl et al.}, 2009;
 Lovekin {\sl et al.}, 2008;  Lovekin \& Goupil, 2009). 
Indeed, unlike $\delta$ Scuti stars,   $\beta$ Cephei  stars 
do not present  significant  outer convective layers
which makes the   mode identification more  trustworthy provided the star is slowly rotating or that its  fast
rotation is taken into account in the mode identification process 
(Ligni\`eres {\sl et al.}, 2006, Reese {\sl et al.}, 2009; Lovekin {\sl et al.} 2008; Lovekin {\sl et al.} 2009).
\begin{figure}[t]
\centering
\epsfig{file=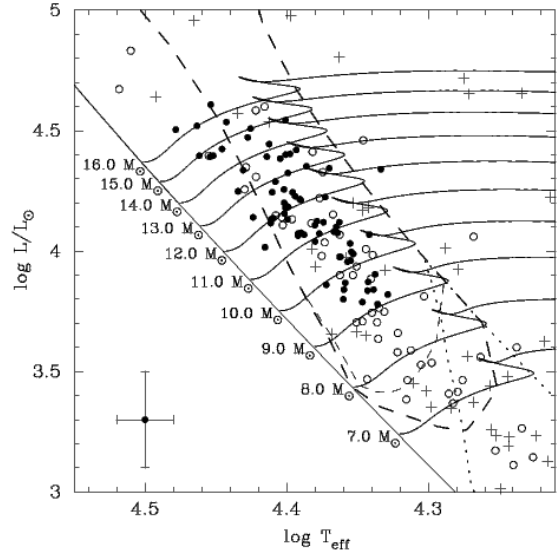,width=8cm}
\caption{{\bf left:} HR diagram  and instability strip for beta Cephei stars. 
Full dots represent confirmed $\beta$ Cephei stars and 
open dots : candidates. The 
dashed lines delimitate the instability strip
for the fundamental radial mode    {\it (adapted from Stankov \& Handler 2005)}.
}
\label{stankov1}
\end{figure}

\subsection{$\beta$ Cephei  stars}
$\beta$ Cephei  stars are main sequence stars with masses 
 roughly larger than $ 5-7 M_\odot$ (Fig. \ref{stankov1}). 
They oscillate with a few low degree, low radial order modes  around the 
fundamental radial mode
i.e. with periods around  3-8 h. 
The modes are excited by the kappa mechanism due to the iron bump opacity. 
These pulsating  stars  
are located at the intersection of the main sequence 
and their instability strip   in a HR diagram (Fig.\ref{stankov1}). For more details about $\beta$ Cephei  stars, we refer to reviews 
by Handler (2006), Stankov \& Handler (2005), Pigulski (2007),  Aerts (2008).

\begin{figure}[t]
\centering
\includegraphics[height=6cm,width=9cm]{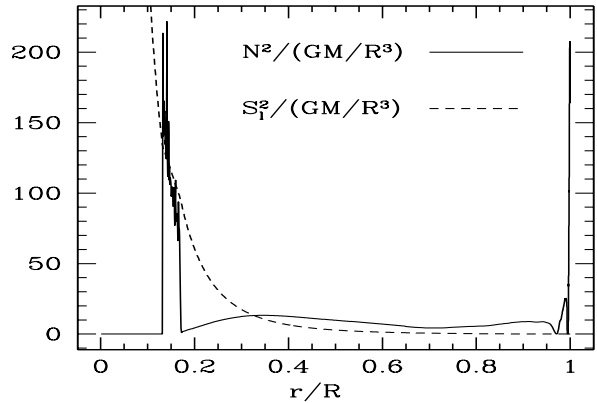}
\caption{ Propagation diagram for 
 model {\bf A}: 
 a 8.5 $M_\odot$ model with $T_{eff}= 22230 K$ 
 and initial $X=0.7,Z=0.019$ (no rotation, no overshoot). The Lamb frequency (dashed line) 
 is plotted for $\ell=1$. 
 Normalized squared frequencies discussed here are found  in the range 
 $\sigma^2$=5-15.
 }
\label{propag}
\end{figure}

Sofar the observed  modes have been  identified  as  p1, p2, g1 modes. 
We recall   that   p  modes are propagative 
when     $ \omega^2 > N^2$  and  $\omega^2 > S_l^2$ (for more details, 
see Christensen-Dalsgaard, 2003 CD03).
The  squared Br\"unt-V\"aiss\"al\"a  (buoyancy) frequency is defined as 
\begin{equation}
N^2 = { g\over r} ~\Bigl({1\over \Gamma_1}  {d\ln p\over d\ln r}- {d\ln\rho\over
d\ln r} \Bigr)
\label{vais}
\end{equation}
with $p,\rho,g$ respectively the pressure,
 density and gravity of the stellar medium and $\Gamma_1$ the adiabatic index.
  The squared   Lamb frequency is defined as 
\begin{equation}
S_l^2 = (k_h c_s)^2= \ell(\ell+1) {c_s^2\over r^2}
\end{equation}
with  $k_h$ the horizontal wavenumber of the pulsation mode 
and $\ell$ the degree of the mode
(when its surface distribution is  described with a spherical 
harmonics $Y_\ell^m(\theta,\phi)$).  
The local sound speed $c_s$   is given by: 
\begin{equation}
c_s = \Bigl({\Gamma_1 p \over \rho}\Bigr)^{1/2}  
\label{cs}
\end{equation}
 For g-modes, the propagative  region is delimitated 
 by  $\omega^2< N^2$  and  $\omega^2 < S_l^2$.

For $\beta$ Cephei stars, mixed modes propagate as  g  mode in the inner part 
and as p mode in the outer part of the star. 
 Depending on the evolutionary stage of the star, one 
expects some of the detected modes to be  of mixed p and g nature. 
Modes with frequencies around that   of the fundamental radial one
 (normalized frequency $\sigma=\omega/\Omega_K \sim 2-3$ with 
 $\Omega_K=(GM/R^3)^{1/2}$,
  $R$ the radius and $M$ the mass of the
 star)  can be mixed modes.
This can be seen in Fig.\ref{propag} which shows 
a propagation diagram for   a typical case, 
 model {\bf A}, a model with  a mass 8.5 $M_\odot$ and an age $=19.9 ~Myr$, 
  a solar metallicity to hydrogen ratio $Z/X=0.019$ with $X=0.7$  and 
  $\log ~T_{eff}=4.347$ and $\log ~L/L_\odot= 3.723 $  
 that  therefore lies  in the middle of the main sequence and instability 
strip for these stars.


 \begin{figure}[t]
\centering
\includegraphics[height=6cm,width=8cm]{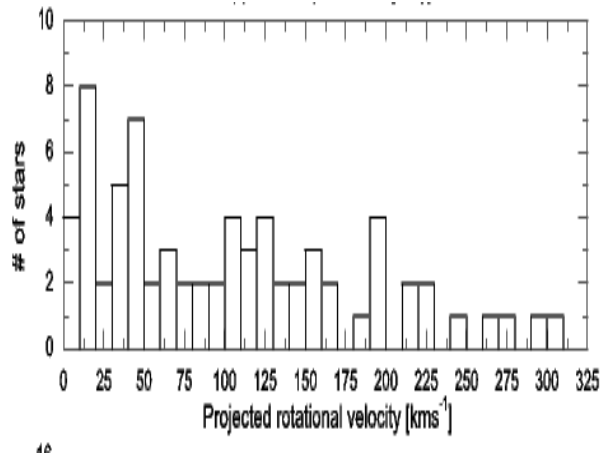}
\caption{ Histogram of projected rotational velocities for $\beta$ Cephei stars
{\it (from Stankov, Handler 2005)}. }
\label{stankov2}
\end{figure}

Rotation of $\beta$ Cephei stars ranges from  
slow (rotational velocity $v<50$ km/s)   to extremely rapid  ($v> 250$ km/s) (Fig.\ref{stankov2}).
Effects of uniform rotation start to modify significantly the tracks
 in a HR diagram beyond $v=100$ km/s
for these masses (Lovekin  \etal, 2009). For $v=100$ km/s, with  a 
stellar radius $R=4.94 R_\odot$, model {\bf A} is characterized by 
  $\Omega/\Omega_K \sim 0.175$ where  
$\Omega_K = (GM/R^3)^{1/2}$ is the break up angular velocity.

\section{ Theoretical framework}

In this section, we  recall  the theoretical framework 
within which seismic observations of these stars can be interpreted in terms of rotation
(for more details, the reader is referred to Goupil (2009) and references therein).
For sake of notation, we recall first  the non rotating case.

\subsection{No rotation} 

Adiabatic pulsation  studies consider the 
linearized conservation equations for a compressible, stratified fluid
about a {\it static equilibrium} stellar  model
 characterized by  $P_0, \rho_0, \Gamma_1, \phi_0$ respectively pressure, density, adiabatic index, gravitational
 potential   profiles. 
 The equation for hydrostatic equilibrium is : 
\begin{equation}
\vec\nabla p_0  = -  \rho_0 ~\vec\nabla \phi_0 
\label{eqhydrost}
 \end{equation}

Assuming  the fluid displacement $\vec {\delta r}(\vec r,t)$ of
 the form $\vec {\delta r}(\vec r,t) =  \vec \xi (\vec r) ~exp(i \omega_0 t) $, 
the linearized  momentum equation then is:
\begin{equation}
{\cal L}_0 (\vec \xi) - \rho_0 ~\omega_0^2 ~\vec \xi = 0
\label{Eqnorot}
\end{equation}
with 
$${\cal L}_0 (\vec \xi)  \equiv \nabla p'+\rho_0 \nabla \phi'+\rho' \nabla \phi_0$$
where ${\cal L}_0$ is a differential  operator acting on  $\vec \xi$;
  $p',\rho',\phi'$ are the Eulerian perturbation for the pressure, density and gravitational potential respectively.
One must add boundary conditions (Unno \etal, 1989) and this gives rise 
to  an eigenvalue problem where  $\omega_0$ is the eigenvalue for the nonrotating case and 
 $\vec \xi$  is the eigenfunction for the fluid displacement. 
In the following, we will keep the notation:
 $\nu$ in $\mu$Hz or c/d ;  $\omega$ in rad/s;   $\sigma= \omega/(GM/R^3)^{1/2}$ 
 the normalized frequency.  One defines the  scalar product:
\begin{equation}
<\vec a| \vec b > 
\equiv \int_V ~ \vec a^* \cdot \vec b ~ {\bf d^3r} 
\label{scalprod}
\end{equation}
where $*$ means complex conjugate and where $V$ is the stellar volume. The scalar product 
of  $\vec \xi$ with Eq.\ref{Eqnorot} then  yields:
$$<\vec \xi|{\cal L}_0 (\vec \xi) - \rho_0  \omega_0^2 \vec \xi > 
\equiv \int_V ~ \vec \xi^* \cdot ({\cal L}_0 (\vec \xi) - \rho_0  \omega_0^2 \vec \xi)~ {\bf d^3r} =0$$
 The 
 eigenfrequency can be obtained as an integral   expression:
\begin{equation}\omega_0^2   ={1\over I} <\vec \xi|{\cal L}_0 (\vec \xi) > \end{equation}
or
\begin{equation}
\omega_0^2   ={1\over I} \int_V ~ \vec \xi^* \cdot (\nabla p'+\rho_0 \nabla \phi'+\rho' \nabla \phi_0) ~ {\bf d^3r}
\label{om2}
\end{equation} 
with
\begin{equation} I = \int_V ~ (\vec \xi^* \cdot \vec \xi) ~\rho_0 ~ {\bf d^3r}
\label{inertia}
\end{equation}

 In absence of rotation, 
the eigenmode displacement  is written in a spherical coordinate system 
with a single harmonics, $Y_\ell^m(\theta,\phi)$,  
with  a spherical degree $\ell$, an azimuthal number
$m$ being the number de nodes along the  equator
\begin{equation}
\vec \xi(\vec r) = \xi_r(r)~Y_\ell^m ~\vec e_r+ \xi_h(r)~\vec \nabla_h Y_\ell^m 
 \end{equation}
where the first part is the radial component and the second term the horizontal
 component of the fluid displacement. The horizontal divergence is 
 $$\vec \nabla_h=  \vec e_\theta~ {\partial \over \partial \theta } +\vec e_\phi ~ {1\over
 \sin \theta }
 {\partial \over \partial \phi} $$

 The divergence of the fluid displacement  is written as:
\begin{equation}
\vec \nabla \cdot \xi = \lambda ~ Y_\ell^m 
 \end{equation} 
with
\begin{equation}
\lambda = {1\over r^2} {d r^2  \xi_r \over dr }- {\Lambda\over r} ~\xi_h
\label{lambda}
\end{equation} 
and $\Lambda= \ell(\ell+1)$. The perturbed   density $\rho' (\vec r)=\rho'(r) Y_\ell^m$ is  given by the linearized continuity equation:
\begin{eqnarray} 
\rho' (r) &=& -\nabla \cdot (\rho_0 \vec \xi)=- {d\rho_0 \over dr}  \xi_r - \rho_0 \lambda 
\label{rhop}
\end{eqnarray}
The  perturbed  gravitational potential $\phi'(\vec r)= \phi'(r) Y_\ell^m$  is given by the perturbed Poisson equation:
\begin{eqnarray}
\nabla^2 \phi' &=&  4 \pi G \rho' 
\label{poisson}
 \end{eqnarray}

The pressure perturbation $p'(r) $ is related to the density perturbation $\rho'(r)$ by the adiabatic
relation (Unno \etal 1989) where $\delta$ means here a Lagrangean variation: 
$$ {\delta p\over p_0} = \Gamma_1 {\delta p\over p_0}$$

\subsection{Including rotation} 
In presence of rotation  
 the  centrifugal and Coriolis accelerations  come into play. 
 The centrifugal force affects the structure of the star - the star is distorted-
 and causes a departure from thermal equilibrium which generates large scale 
 meridional circulation and  chemical mixing. 
Accordingly, the resonant cavities of the modes are modified.
  The static equilibrium  (averaged over horizontal surfaces)  1D stellar  model 
  is modified and characterized by 
$P_{0,\Omega}(r), \rho_{0,\Omega}(r), \Gamma_{1,\Omega}(r), \phi_{0,\Omega}(r)$ 
with $\Omega(r,\theta)$ the  rotation rate.
The Coriolis force enters the equation of motion 
and   affects the motion of waves and frequencies of normal modes. 
The linearized equation of motion  is modified. As rotation  breaks the azimuthal symmetry, it 
lifts the frequency degeneracy:  
without rotation, $2\ell+1$ modes with given $n, \ell, m$  
($m=-\ell,\ell$)  have  the same frequency $\omega_{0}$ (omitting for shortness the subscripts 
$n,\ell$). With rotation, 
the same  $2\ell+1$ modes have different frequencies 
 $\omega_{m}$ and the rotational splitting
 is defined as :  $S_m= (\omega_{m}-\omega_{0})/(m)$.  One also uses
  $S_m= \omega_{m}-\omega_{m-1}$ and
the generalized rotational splitting: 
\begin{equation}
S_m = {\omega_{m}-\omega_{-m} \over 2 m}
\label{Sm}
\end{equation}
where $\omega_{m}$
is the mode frequency.
These various definitions  are equivalent only at first
 perturbation order in the rotation rate $\Omega$; the  first two
  are used when only a 
 few components are available.  


\subsection{ Rotational splittings}\label{rotspl}
At first perturbation order in $\Omega$, 
only the Coriolis acceleration plays a role.
The linearized equation of motion including
 the effect of Coriolis acceleration ($2\Omega \times \vec v$)
  in a frame of inertia
 is 
\begin{equation}
{\cal L}_0 (\vec \xi) - \rho_0 ~(\omega_m+m\Omega)^2 \vec \xi  +2 \rho_0 ~(\omega_m+m\Omega)~ \Omega ~{\cal K} \vec \xi= 0
\label{eqCor}
\end{equation}
with $  {\cal K} \vec \xi =   i \vec e_z \times \vec \xi$ and 
$\vec \xi$  is the displacement
eigenvector in absence of
rotation and $e_z$ is the  vectical unit vector in cylindrical coordinates. 
The nonrotating case is recovered by setting $\Omega=0$.
One then expands the displacement eigenfunction 
as $\vec\xi =\vec\xi_0+\vec\xi_1$ and the eigenfrezquency 
as $\omega_m=\omega_{\Omega=0}+\omega_{1,m}$ 
where $\omega_{\Omega=0}$, $\vec\xi_0$ correspond to the eigenfrequency and eigenfunction 
for a nonrotating star and 
$\omega_{1,m}$, $\vec\xi_1$ give the first order 
correction due to Coriolis acceleration. Keeping only
terms up to $O(\Omega)$, one obtains:
\begin{eqnarray} 
 {\cal L}_0(\vec \xi_1) &-& \rho_0 ~\omega_{\Omega=0}^2 ~\vec \xi_1- 2 \rho_0 ~ \omega_{\Omega=0}(\omega_{1,m}  +m \Omega) ~\vec \xi_0 \nonumber \\  
 &+& 2 \rho_0 ~\omega_{\Omega=0} ~\Omega ~{\cal K} \vec \xi_0 =0 
\label{sp0}
\end{eqnarray}
The correction to the eigenfunction $ \vec \xi_1$ can be chosen so that $<\vec \xi_0|\vec \xi_1>=0$.
Taking the scalar product Eq.\ref{scalprod} of $\vec \xi_0$ with Eq.\ref{sp0} and keeping only
terms up to $O(\Omega)$ yields: 
\begin{eqnarray} <\vec \xi_0| \left[ \right. {\cal L}_0(\vec \xi_1) &-& \rho_0 ~\omega_{\Omega=0}^2 ~\vec \xi_1- 
2 \rho_0 ~\omega_{\Omega=0} ~(\omega_{1,m}  +m \Omega) ~\vec \xi_0 \nonumber \\  
 &+&2 \rho_0 ~\omega_{\Omega=0} ~\Omega ~{\cal K} \vec \xi_0 \left. \right]>=0 
\end{eqnarray}
from which one derives  for a mode with given $n,\ell$ 
\begin{eqnarray}  
 ~\omega_{1,m} ~I_0 &=&\int_V ~ \vec \xi_0^* \cdot  (\Omega ~{\cal K} -m \Omega 
) ~ \vec \xi_0~ \rho_0  ~ {\bf d^3r}
\end{eqnarray}
which is rewritten as:
\begin{equation}
 \omega_{1,m} =    \int_0^R \int_0^\pi  K_{m}(r,\theta) ~{\Omega(r,\theta)} ~  d\theta dr
\label{sp}
\end{equation}
where the analytical expression   for the kernels 
$K_{m}$ is given in Appendix. 
At first order $O(\Omega)$, the generalized splitting Eq.\ref{Sm} 
then is given by 
\begin{equation}
S_m= {\omega_{1,m}-\omega_{1,-m} \over 2m}
 \label{om1m}
 \end{equation}
  Assuming a shellular rotation $\Omega(r)$, the splitting $S_m$ becomes  $m$ independent and one has:
\begin{eqnarray} 
S   =   ~\int_0^R  ~K(r) ~{\Omega(r)} ~dr 
\label{sp4}
\end{eqnarray}
with the rotational kernel 
 \begin{equation}
 K(r) =-{1\over I}  \left( \xi_r^2  -2 \xi_r \xi_h + (\Lambda -1) \xi_h^2\right) ~\rho_{0}~ r^2
 \label{Kr}
 \end{equation} 
and mode inertia (Eq.\ref{inertia}):
\begin{equation}
I=\int_0^R   ~(\xi_r^2 +\Lambda \xi_h^2) ~\rho_{0} r^2 dr
\end{equation}  
with again $\Lambda = \ell(\ell+1)$ and $R$ the stellar radius.
For a uniform rotation, this  further simplifies to   
\begin{equation}
S= \Omega ~\beta  ~~~~~~~~;~~~~\beta= \int_0^R  ~K(r)  ~dr 
\label{beta}
\end{equation}
$\beta$    is assumed to be  known from an 
appropriate stellar model, $S$ is measured and $\Omega$ is inferred. This 
 will be used in Sect.4 for $\beta$ Cep stars.

When only a few measured splittings are available, information about the internal rotation is
limited so one  assumes for instance a uniform rotation for the convective 
core with the angular velocity $\Omega=\Omega_c $ (for $x=r/R \le x_c$) and a uniform
rotation for the envelope $\Omega=\Omega_e$ for $x> x_c$. 
Both values are the unknowns. Inserting into Eq.\ref{sp4}, 
$$S= \int_0^{1} K(x) ~\Omega(x) ~dx =\Omega_c \beta_c+\Omega_e \beta_e $$
with 
$$ \beta_c = \int_0^{x_c} K(x) ~dx ~~;~~ \beta_e = \int_{x_c}^1 K(x) ~dx$$ 
The detection  of 2 triplets $\ell=1$ for instance yields 
$\Omega_c$, $\Omega_e$  and $\Omega_c/\Omega_e$ 
provided $ \beta_c,\beta_e $ are given by a model close to the observed star. 
This type of approach  was used  to determine whether the star is in rigid rotation or not
 for a $\delta$ Scuti star (Goupil {\sl et al.}, 1993); for 
white dwarfs (Winget {\sl et al.} 1994; Kawaler {\sl et al.}, 1999)
 and  recently  for $\beta$ Cephei stars  (Sect.4 below)
  and SdB stars (Charpinet \etal, 2008).

\subsection{Splitting asymmetries: distorsion}\label{asym}
At second order in  the rotation rate, 
the centrifugal acceleration comes into play. This has several consequences on the oscillation
frequencies (for a review  Goupil,  2009). One is that the split components are  no longer equally spaced.
It is then convenient to define $A_m$ the splitting asymmetry 
as
 \begin{equation}
 A_m  = \omega_0 -{1\over 2}(\omega_m+\omega_{-m})
 \label{Am}
 \end{equation} 
In order to interpret observed asymmetries, let consider a given multiplet of modes
 (i.e. with specified $n,\ell$). Its oscillation
frequencies, $\omega_m$ ($m=-\ell,..,\ell$), are computed up to second order $O(\Omega^2)$ as:
  \begin{equation}
  \omega_m =\omega_{0,\Omega} + m ~ S_|m|+ {{\bar \Omega}^2\over \omega_{0,\Omega}} ~(X_1 +m^2 X_2) 
  \label{omeg2}
 \end{equation} 
whereand $\omega_{0,\Omega} $ is the eigenfrequency for a static model including the horizontally averaged 
centrifugal acceleration. The second term is the  splitting Eq.\ref{om1m} due to Coriolis effect
 and 
$ {\bar \Omega}$  is an averaged rotation rate. 
The last term is the asymmetry due to the non spherical part of the centrifugal distorsion  
which dominates   for high radial order modes.  Expressions for $X_1,X_2$ can be found in Saio (1981), DG92, Soufi {\sl et al.} (1998); 
Suarez \etal (2006); Goupil  (2009).
For low radial modes such as those excited in $\beta$ Cep
stars, the second order Coriolis contributions to $X_1,~X_2$ remain significant. 
 According to Eq.\ref{omeg2}, the asymmetry is then  given by:  
\begin{equation}
A_m =  \left({    {\bar \Omega}^2 \over \omega_{0,\Omega}}\right) ~m^2~X_2 
\end{equation} 
 Let 
 consider again the linearized equation of motion   including  now  the centrifugal
acceleration: 
\begin{equation}
{\cal L}_{0,\Omega} (\vec \xi) - \rho_{0,\Omega}~ \hat \omega^2 ~\vec \xi  +2 \rho_{0,\Omega} \hat \omega \Omega K
(\vec \xi) + {\cal L}_2 (\vec \xi) -\rho_2 ~\hat \omega^2 ~\vec \xi= 0
\label{eqcent}
\end{equation}
where $\hat \omega= \omega_m+m  \Omega$. 
The spherical part of the centrifugal acceleration is included in  the spherical  1D model, therefore the linear
operator depends on the rotation rate i.e.
\begin{equation}
{\cal L}_{0,\Omega} (\vec \xi)  \equiv \nabla p'+\rho_{0,\Omega} \vec \nabla \phi'+\rho'  \vec \nabla \phi_{0,\Omega}
\end{equation}
and   for the non spherical distorsion
\begin{equation}
{\cal L}_2 (\vec \xi)= \rho' ({\rho_2 \over \rho_{0,\Omega}} \nabla p_{0,\Omega}-\nabla p_2) + \rho_2 \nabla
\phi' + \rho_{0,\Omega}  ~\vec e_s ~ r \sin
\theta  ~\vec \nabla \Omega^2 \cdot \vec \xi
\end{equation} 
where $ \vec e_s = \sin \theta~  \vec e_r + \cos \theta ~ \vec e_\theta$  
in a spherical coordinate
system ($\vec e_r, \vec e_\theta, \vec e_\phi $).The  subscript 2 indicates  departure from sphericity $p_2,\rho_2,\phi_2$ for the pressure, density and 
graviational potential respectively. 
Again using the scalar product Eq.\ref{scalprod}, 
one writes
\begin{eqnarray}
<\vec \xi_0|{\cal L}_{0,\Omega} (\vec \xi) &-& \rho_{0,\Omega}  ~\hat \omega^2 ~\vec \xi 
 +2 \rho_{0,\Omega} ~ \hat \omega ~  \Omega ~  K(\vec \xi)>  \nonumber \\ 
& + &<\vec \xi|({\cal L}_2 (\vec \xi) -\rho_2 \hat \omega^2 ~\vec \xi)>= 0
\label{second}
\end{eqnarray}

One then assumes an eigenfunction of the form $\vec \xi=\vec \xi_0+\vec\xi_1+\vec \xi_2$ and the eigenfrequency 
as $\omega_m=\omega_{0,\Omega}+\omega_{1m}+\omega_{2}$ where the unknown 
now is $\omega_2$.
Solving Eq.\ref{second} for $\omega_{2m}$ leads to an integral expression 
for $X_1,X_2$ and therefore an expression for 
 $A_m$   of the form:
\begin{equation}A_m =m^2  \, \int_0^1  ~ \Omega^2(x) ~ K_{2}(x)  ~dx
\label{Am}
\end{equation} 
where 
$K_2(x)$ include effects of distorsion of the structure throughout  $p_2,\rho_2$ and depend on the 
 eigenfunction $\vec \xi$.  
 Detailed expression for $K^{(j)}_2(x)$  can be found  in DG92, Soufi {\sl et al.} (1998), 
 Karami (2008), Suarez {\sl et al.} (2006), Goupil (2009). An exemple for $K_{2}(x)$ is shown in Fig.\ref{asym} and discussed
 in Sect.4.2

Splitting asymmetries  can provide  probes of the internal structure
 which differ from those  derived with  the splittings $S_m$
 as  the corresponding kernels are different. When only a few  observed asymmetries are available,
 one can proceed  as for the  splittings (Sect.\ref{rotspl} above).  Assuming a rotation profile  of the simplified form:
\begin{eqnarray}
\Omega^2(x) &=& \Omega_c^2 ~~ {\rm for}  ~~ x_c < x  \nonumber  \\
\Omega^2(x) &=& \Omega_c^2 + 2 (x-x_c) ~\Omega' \Omega_c + (x-x_c)^2 ~\Omega^{'2} ~~ {\rm for}  ~~   x_c \le x \le x_e \nonumber  \\
\Omega^2(x) &=& \Omega_e^2  ~~ {\rm for}  ~~ x_e < x 
\label{omegs}
\end{eqnarray}
with
\begin{equation}
\Omega'= {\Omega_e-\Omega_c \over x_e-x_c}
\end{equation}
then 
\begin{equation}
A_m = m^2 \Bigl( \Omega_c^2  ~\beta_{2,0} + 2 \Omega' \Omega_c ~\beta_{2,1} + \beta_{2,2}
~\Omega^{'2} +\beta_{2,e} \Omega_e^2 \Bigr)
\end{equation}
where 
$ \Omega_c,\Omega'$  are assumed known from the splittings (Sect.3.3) and
\begin{eqnarray}
\beta_{2,q} &=& \int_{x_c}^{x_e}  ~ (x-x_c)^q ~K_{2}(x)~ dx \\
\beta_{2,e} &=& \int_{x_e}^{1}  ~ K_{2}(x)~ dx 
\label{betas}
\end{eqnarray}

Determination of  the $\beta_2$ coefficients then   brings some information on  the kernels 
$K_{2}(x)$ with the promising prospect of deriving constrains 
 on  the rotationally distorted part of the stellar structure.

\subsection{Axisymmetric modes: rotationally induced mixing}

Centrifugal departure   from spherical symmetry  has important  effects on  all modes including the axisymmetric modes. 
Indeed the values of the $m=0$ mode frequencies  are  shifted when compared to those of non rotating models.
 Hence  the differences 
 \begin{equation}
 \delta \omega= \omega_0 - \omega_(\Omega=0) = \left({ {\bar \Omega}^2 \over \omega_{0,\Omega}}\right) ~m^2~X_1
 \end{equation} 
 from Eq.\ref{omeg2} 
 between frequencies of a given mode from a model including rotation and
  a non rotating model can be an efficient diagnostic for rotation effects
 although   some care  must be taken when  defining the $\Omega=0$ stellar
  model for comparison.  
This has been extensively
discussed in past  publications (Chandrasekhar \& Lebovitz, 1962; Saio,
 1981; Gough \& Thompson, 1990; DG92; CD03, for a review, see Goupil, 2009). 

Another (indirect)  effect   of the star oblatness  on frequencies, 
as already mentionned in Sect.\ref{MMS},  is  due to the departure from   radiative equilibrium  which generates
 large scale motions (meridional circulation),
differential rotation and consequently shear turbulence. 
All this concurs to affect the rotation profile. It also
causes mixing of chemical elements which affects the prior evolution of the observed star and therefore 
its structure. These structure changes must be computed by coupling both evolutions of
the angular momentum and the chemicals, as already mentionned in Sect.2.  
 These equilibrium structure  modifications  affect all modes as compared to those of a nonrotating star, 
including the axisymmetric modes. 
The effect on the frequencies can be quite significant as was discussed by
 Goupil \& Talon (2002) and
quantified by Montalban {\sl et al.} (2008), Miglio {\sl et al.} (2008),
 Goupil \& Talon (2008)  (see Sect.4.3 below)

We consider here  only the effect of  the structure modifications due  to rotationally induced mixing  on the axisymmetric mode 
frequencies. The Coriolis or the centrifugal accelerations then are not included in the linearized  equation of motion.
 Hence  the linearized equation of motion including rotationally induced mixing 
  yields the usual    integral expression for the eigenfrequency of a nonrotating model, Eq.\ref{om2},
except for  the structure quantities such as the density, the pressure, the gravity 
($\rho, p, g$ resp. ) etc... which  are 
 modified by  the rotationnally induced mixing.  
 As they now depend on the rotation rate,
  we write them as $\rho_{\Omega}, p_{\Omega}, g_\Omega ...$. 
  The linearized equation of motion including rotationally induced
   mixing in a 1D spherically symmetric stellar model  then is given by:
\begin{equation}
\omega_{0,\Omega}^2 ={1\over I_\Omega} \int_V ~ \vec \xi_\Omega^* 
\cdot (\vec \nabla p'+\rho_\Omega \vec \nabla \phi'+\rho' \vec \nabla
\phi_\Omega ) ~ {\bf d^3r} 
\label{eqcent2}
\end{equation}
with the mode inertia:
$$I_\Omega = \int_V ~ (\vec \xi_\Omega^* \cdot  \vec \xi_\Omega)~ \rho_\Omega ~{\bf d^3r} $$
From now on for sake of shortness,  
 we omit the subscript $\Omega$ for the eigenfunctions.
We define the dimensionless variables according to Dziembowski (1971)
 (see also Unno {\sl et al.}, 1989):
\begin{eqnarray}
y_1 &=& {\xi_r \over r}; ~~ y_2= {1\over g_\Omega r}~(\phi'+{p'\over \rho_\Omega}) ; ~~y_3={1\over g_\Omega r}~\phi' 
\end{eqnarray}

Starting with Eq.\ref{eqcent2}, integrations over surface angles and  a few integrations by part   for the
radial part yield:
\begin{equation}
\omega_{0,\Omega}^2 ={1\over I_\Omega} \int_R ~ (- \lambda (y_1+y_2) ~ 
-{d\ln \rho_\Omega \over d\ln r} ~ y_1~(y_1+y_3))~ g_\Omega~ \rho_\Omega ~r^3 ~dr
\label{eqcent3}
\end{equation}
where we have assumed that the surface integrals vanish. From its definition
Eq.\ref{lambda}, 
$$ \lambda =V_{g,\Omega}~ (y_1-y_2+y_3) $$
with $$V_{g,\Omega} = - {1\over \Gamma_{1,\Omega}}~ {d\ln p_\Omega \over d\ln r} $$
.
 Note that there are several alternative equivalent expressions for  $\omega_{0,\Omega}^2$.

 Differences  between the structure 
of a model including rotationnaly induced mixing 
and that  of a  model which does
not   result in differences in the eigenfrequencies
which we note $\delta \omega = \omega_{0,\Omega}-\omega_{0,\Omega=0} $. 
We will see in Sect.4.3 that the structures of  the models indicate 
that $p_\Omega$  and its derivative with
respect to the radius,
 the gravity $g_\Omega$, the density $\rho_\Omega$  
are  not significantly modified compared to the derivative of the density with
respect to the radius.
Accordingly using Eq.\ref{eqcent3}  and keeping only the first order terms, one
obtains:
\begin{eqnarray}
\delta \omega &\approx & - {1\over 2~ \omega_{\Omega=0} ~I_{\Omega=0} } \times \nonumber \\
& & \int_R ~ 
\delta({d\ln \rho_\Omega \over d\ln r}) ~
 y_1~(y_1+y_3)~ g_{\small\Omega=0}~ \rho_{\Omega=0}~ r^3 dr
\label{eqcent4}
\end{eqnarray}
where we have also assumed that
the perturbations of the eigenfunctions $y_{j,\Omega}-y_{j,\Omega=0}$ ($j=1,3$) 
are negligible at first order. 
For massive main sequence stars the largest difference 
 $\delta({d\ln \rho_\Omega/ d\ln r}) $ arises near the convective core  (Sect.4.3). Largest 
frequency differences therefore are expected for  mixed modes compared to p-modes.
Note that the same interpretation can be obtained with  differences in the
Br\"unt-V\"aiss\"al\"a behavior. Indeed  at the same level of approximation,
 one has from Eq.\ref{vais}
 \begin{equation}
 \delta \Bigl( {r N^2 \over g} \Bigr) = -\delta   \Bigl({d\ln \rho_\Omega \over d\ln r}\Bigr)
\end{equation}

For high frequency (i.e.  pure) p-mode  which propagate significantly above
 the $\vec \nabla \mu $ region, the
difference  $\delta   \Bigl({d\ln \rho_\Omega / d\ln r}\Bigr)$ is essentially negative. 
In addition $|y_3| <<| y_1|$ so that we obtain 
\begin{eqnarray}
\delta \omega &\approx & - {1\over 2~ \omega_{\Omega=0} ~I_{\Omega=0} } \times \nonumber \\
& & \int_R ~ 
\delta({d\ln \rho_\Omega \over d\ln r}) ~
 y^2_1 ~ g_{\Omega=0}~ \rho_{\Omega=0}~ r^3 dr ~~~ >0
\label{eqcent5}
\end{eqnarray}
which is small and positive. For mixed modes having high amplitude in the $\vec \nabla \mu $ region, 
 $\delta   \Bigl({d\ln \rho_\Omega/ d\ln r}\Bigr)$  can be positive  and the  frequency
 difference can be large and  negative as illustrated in Sect.4.3.
The difference $\delta \omega$ is quantified and discussed  in the case of 
 a $\beta$ Cephei model in Sect.4.3.

\section{Seismic analyses of  four $\beta$ Cephei stars}

\begin{figure}[t]
\includegraphics[height=6cm,width=6cm]{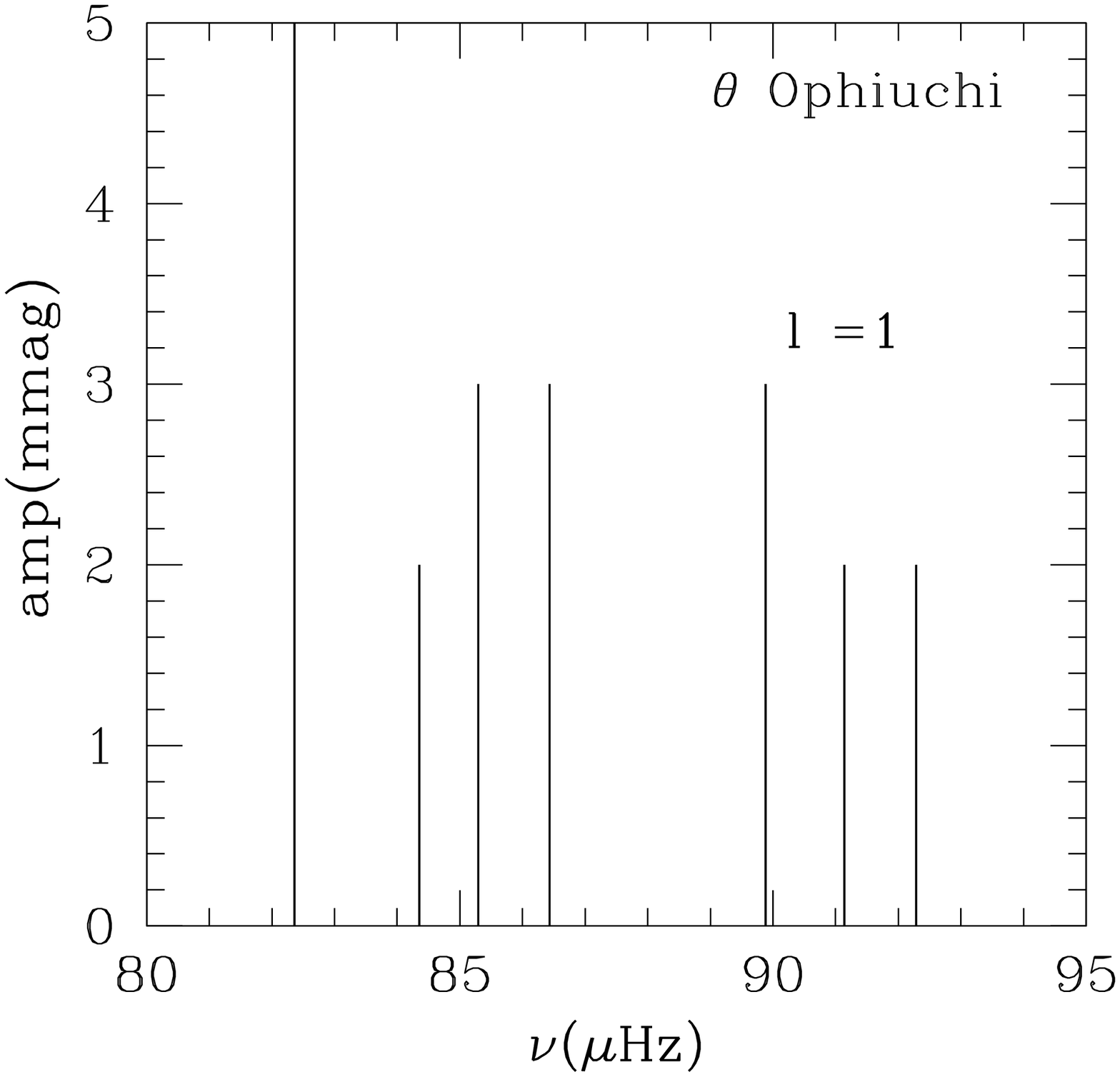}
\includegraphics[height=6cm,width=6cm]{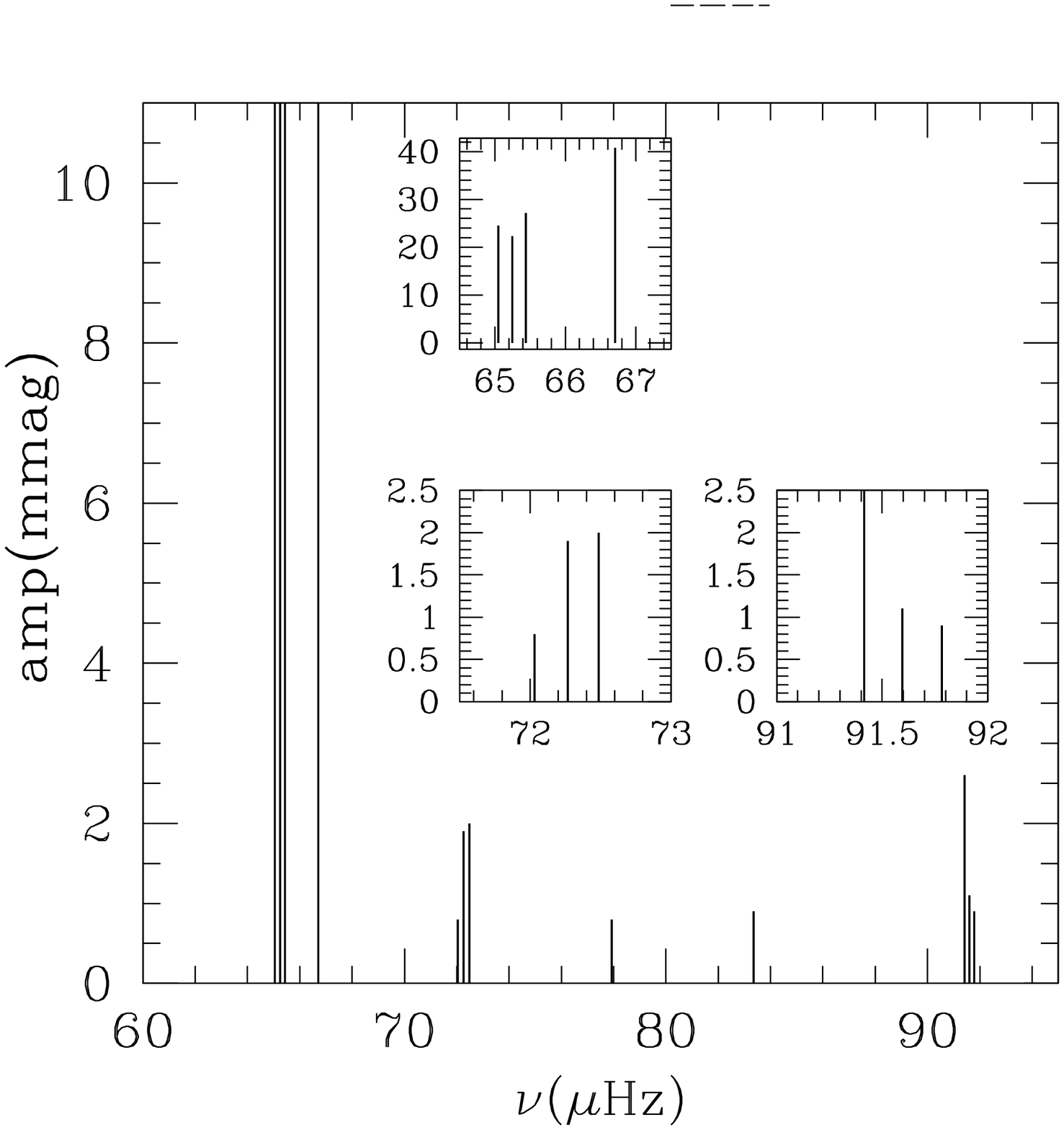}
\caption{Schematic representation of the  
power spectrum of {\bf left:}  $\theta$ Ophiuchi {\it (data from Handler \etal, 2005)} 
 and {\bf right:} 
$\nu$ Eri  {\it (data from Jerzykiewicz \etal, 2005)}}
\label{nueri}
\end{figure}

We discuss 4 $\beta$ Cephei stars which  have been the subject of seismic analyses and 
for which information about rotation and core overshoot has been inferred:
V836 Cen (HD 129929); $\nu$ Eridani,  $\theta$ Ophiuchi and 12 Lacertae (see also Thoul, 2009). 
Schematic representations of
the frequency spectra for the first three stars are displayed in Fig.\ref{HD129} and
Fig.\ref{nueri}. These four stars are relatively slow rotators (with surface rotaétional
velocities smaller than $\approx$ 70 km/s). Determination of the luminosity, 
effective temperature and location in the HR diagram for these slow rotators are not
significantly affected by rotation.

\subsection{Rotational splittings}

\subsubsection{HD129929}  
\begin{figure}[t]
\includegraphics[height=6cm,width=6cm]{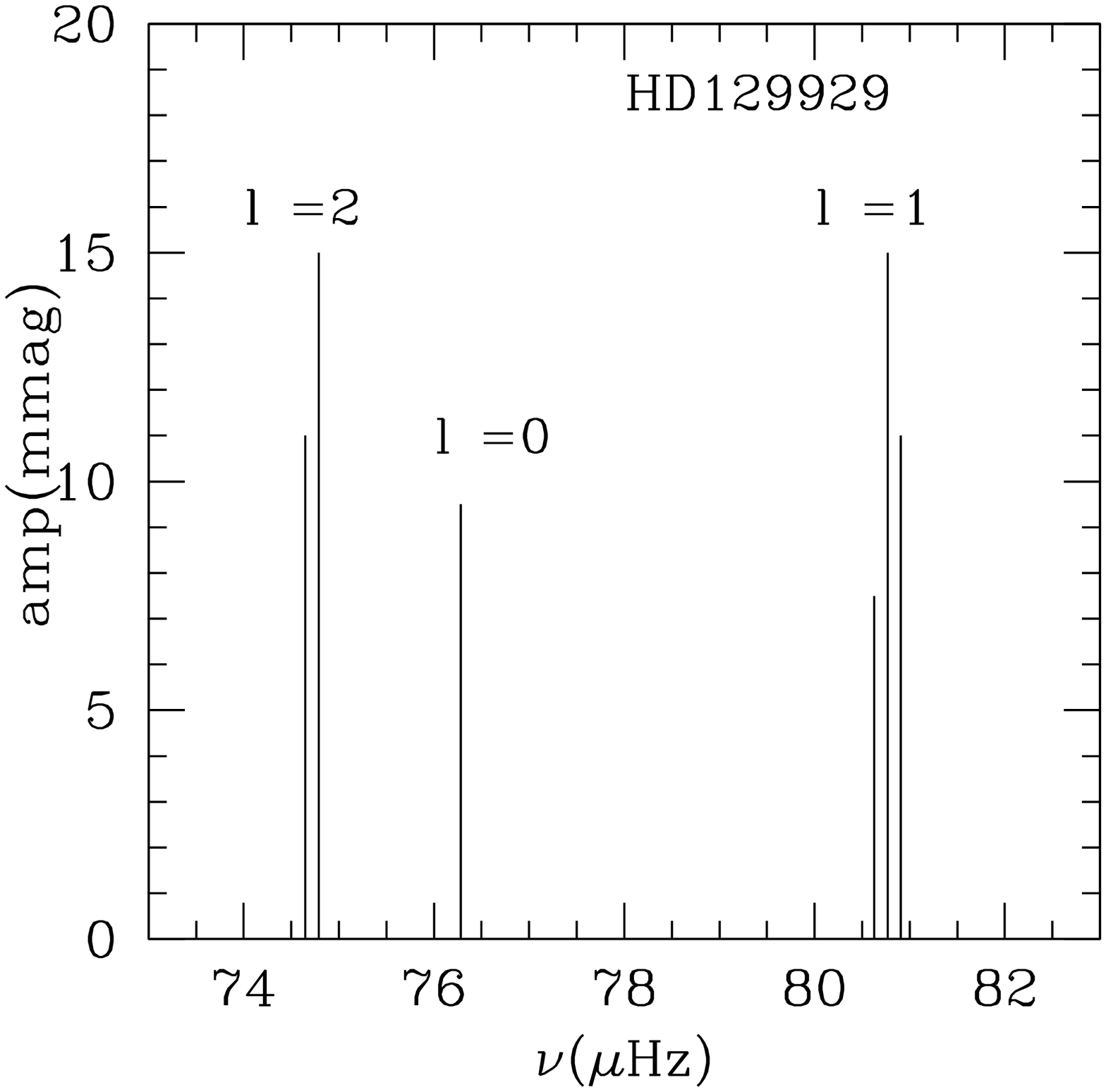}
{\vskip -6 cm }{\hskip 6.truecm}
\includegraphics[height=5.5cm,width=6cm]{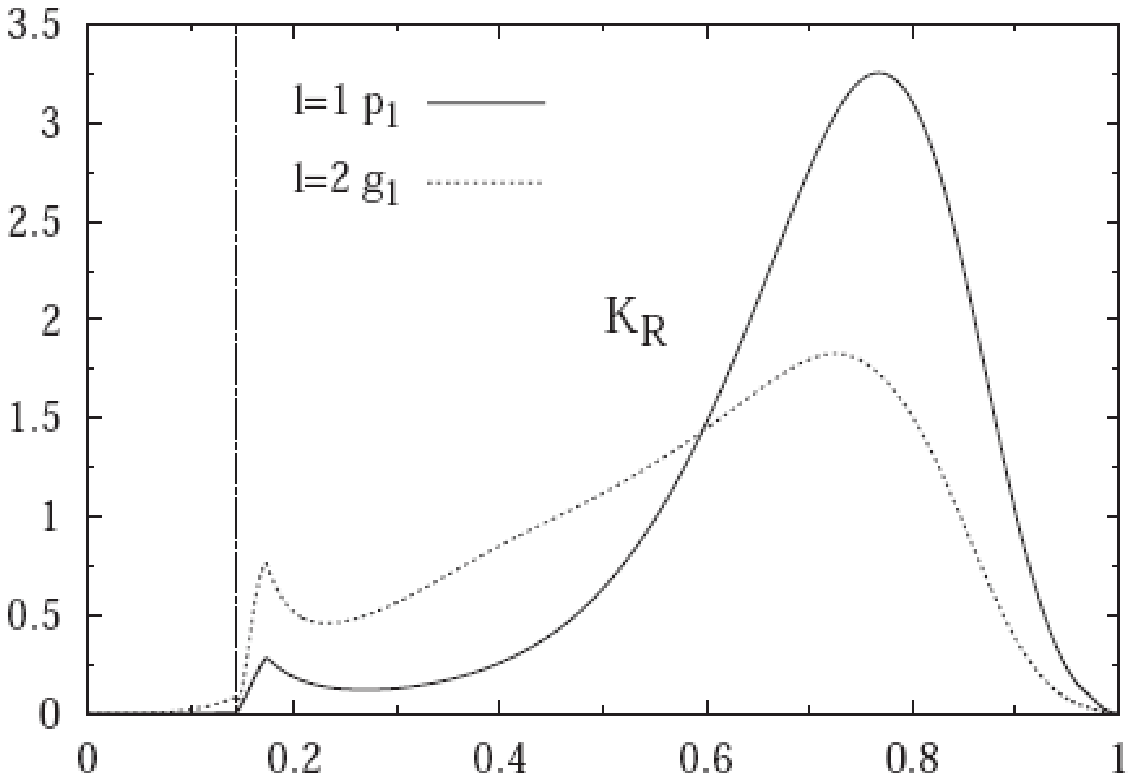}
 {\vskip 0.5 cm }
\caption{{\bf left:} Schematic representation of the  
frequency spectrum of  HD129929 {(\it data from Aerts \etal, 2004)}.
{\bf right:} Rotational kernels for the  excited  p1 and g1 modes   of HD129929 in function of the radius 
$r/R$ normalized to the stellar radius {\it (from Dupret \etal, 2004)}.}
\label{HD129}
\end{figure}

is a  main  sequence $\sim 9 M_\odot$ star  
for which one $\ell=1,~ p_1$
 triplet has been detected and identified  as well as one radial mode 
 and  2 successive components 
 of the $\ell=2, ~g_1$ mode as represented in Fig.\ref{HD129} (Aerts {\sl et al.}, 2004, Dupret {\sl et al.}, 2004). 
From the triplet and assuming a solid body rotation, 
, one uses  $S= \Omega~\beta$ (Eq.\ref{beta}) as explained in Sect.\ref{rotspl}.
 With $\beta$ known  from an appropriate stellar  model,  
 the  measured splitting for the $\ell=1, p=1$ triplet $S$  gives 
 $v_{rot}= 3.61$\,km/s  
but from the two successive components of the $\ell=2$ multiplet, 
 one obtains $v_{rot}=4.21$\,km/s, 
clearly indicating  a nonuniform rotation (Dupret et al., 2004).  
Assuming therefore a uniform rotation for the convective core with angular velocity $\Omega=\Omega_c$ and
a uniform rotation $\Omega=\Omega_e$ for the envelope of the star, 
the splittings then obey  $S = \Omega_c \beta_c +\Omega_e \beta_e $ where $\beta_j$ are 
the integral for the core or  the
envelope (Sect.\ref{rotspl}). It is found that $|\beta_c| << |\beta_e|$ 
that-is actually  the detected 
 modes do not efficiently probe the convective core. This can be seen
  with the associated rotational kernels  in Fig.\ref{HD129} which have no amplitude in the core.  Therefore $\Omega_c$ is  
 taken as the rotation rate of the
 radiative  region  in the $\mu$-gradient region  above the convective core (with $\mu$ the mean molecular
 weight). 
 Assuming a linear depth variation of the angular velocity in the envelope $\Omega(x) = \Omega_0 +(x-x_0) \Omega_1$,
the splittings must obey $S= \Omega_0 \beta_0+\Omega_1\beta_1 $ where again
$\beta_0$ and $\beta_1$ are known from the stellar model; 
$$ \beta_0 = \int_0^{x_c} K(x) ~dx ~~;~~ \beta_e = \int_{x_c}^{x_e}~  K(x) (x-x_c)~dx$$
The knowledge of $S_1$ and $S_2$
then yields $\Omega_0$  and $\Omega_1$.  
A  rotation gradient in the envelope with $\Omega_c/\Omega_e=3.6$  is 
  obtained. 
  
In addition,   the seismic modelling of the detected axisymmetric modes
 favors a core overshooting distance of
   $\sim 0.1$  pressure scale height ($H_p$) rather than 0 while
  an overshoot of $0.2\,H_p$ is rejected.

\subsubsection{$\theta$ Ophiuchi} is also 
 a main sequence  $\sim 9\,M_\odot$ star  with 
 an effective temperature $T_{eff}\sim 22 900 ~K$.
 Three multisite campains 
 seismic observations and data analyses   reveal 
7  identified frequencies: the radial fundamental $\ell=0$ (p1);  one  triplet $\ell=1 (p1)$ and  
  3 components $(m=-1,1,2)$ of a  quintuplet $\ell=2$ (g1) (Handler \etal,  2005). 
A seismic analysis  led  Briquet {\sl et al.} (2007) to conclude that  the case of $\theta$ Ophiuchi is  similar to HD129929. 
The detected  modes do not provide strong constraint about the rotation of the convective core. 
On the other hand, unlike HD129929, the data for $\theta$ Ophiuchi
 are compatible with a uniform or a quite slowly varying rotation of the   envelope.
The convective core overshoot distance is found to be $(0.44\pm 0.07) ~H_p $ This is a much larger
amount than found for HD129929. Whether this difference must be related to the fact that 
$\theta$ Ophiuchi seems to rotate more than 10 times faster than HD129929
remains an open issue.


\subsubsection{$\nu$ Eri} is a very interesting case as it 
oscillates with 3 triplets $\ell=1$ ($g_1$, $p_1$, $p_2$), one radial mode $p_1$
and one $\ell=2$ component.   Seismic studies 
 show that  the detected modes are able to probe the rotation of the core, which is rotating faster 
     than  the envelope (Pamyatnykh {\sl et al.}, 2004, PHD04;  Ausseloos {\sl et al.}, 2004; 
     Dziembowski \& Pamyatnykh, 2008 (DP08); Suarez {\sl et al.} 2009).
DP08   further   assumed  a linear gradient as a transition (in the $\mu$ gradient zone) 
     between the uniform fast
    rotation $\Omega =\Omega_c$  of the core and the uniform slow rotation of 
    the envelope $\Omega=\Omega_e$ above the $\mu$-gradient region.  
     They find   a ratio $\Omega_c/\Omega_e=5.3-5.8$. 
         Model fitting  based on the 3  axisymmetric $\ell=1$ modes 
      yield an extension of the mixed central region
of 0.1-0.28 $H_p$ above the convective core  radius depending on the adopted
chemical mixture and metallicity value (DP08; Suarez {\sl et al.}, 2009).

\begin{table}
\caption{Overshoot versus rotation rate for several stars from seismic analysis. $Veq$ is the derived  the equatorial
velocity,  $ \alpha_{ov}$ the overshoot parameter, $\Omega_{inner}/\Omega_{env}$ the ratio of the
 the rotation rate in the inner layers to that of the surface,  $Z$ the metallicity. The modellings assume a
 Grevesse-Noels mixture except for $12 Lac $.
\label{coef_fit}}
\begin{tabular}{lccccc}
\hline
$\beta$ Cep & Veq (km/s) & $ \alpha_{ov}$ & $\Omega_{inner}/\Omega_{env}$  & Z  & ref\\
\hline \hline
HD129929   & $\sim$ 2  &  0.1  $\pm$ 0.05 & $ \Omega_{0.2}/\Omega_{surf} \sim3.1$   &  0.019 $ \pm$ 0.003 &(1)\\
$\theta$ Ophiuchi & 29  $\pm$ 7 &  $0.44^* \pm$ 0.07 & env. unif. rotation  &  0.012 $\pm $0.003  &(2)\\
$\nu$ Eri  & $\sim$ 6   & 0.15 $\pm$  0.05  & $\Omega_ {c}/\Omega_{env} \sim 5.5-5.8 $  & 0.0172 $ \pm$0.0013  &(3) \\
$12 Lac $ & $\sim$ 49&  $ <0.4$  &$\Omega_{c}/\Omega_{env}  \sim  1.8-5 $ &0.01-0.015    & (4), (5) \\
\hline \hline
*Asplund mixture  \\
\end{tabular}
(1) Dupret {\sl et al.}, 2004 (2) Briquet {\sl et al.}, 2007 (3) Pamyathnyck {\sl et al.}, 2004, (4) DP08 (5) Desmet {\sl et al.},  2009
\end{table}

\subsubsection{12 Lac}
       
Several frequencies have been detected for this star (Handler \etal,  2006) but 
   only 4 of the detected frequencies correspond to identified $(\ell,m)$
   modes (Desmet \etal,  2009).  Only 2 successive 
   components of one $\ell=1$ triplet  are known
   which is not enough to  provide information on the  inner/surface rotation
   ratio.  One can use as an  additional information    the equatorial 
    surface value,  $v_{eq}= 49\pm 3$ km/s as derived by
   Desmet {\sl et al.} (2009). One needs the stellar radius which is 
   derived from a seismic modelling of the star. The resulting 
   seismic model and its radius  depend on 
   the radial orders  identified for the modes 
   (DP08 and  Desmet {\sl et al.}, 2009). 
   Second order (centrifugal) effects on the frequencies must also be taken into account
   as the rotation for 12 Lac seems to be fast enough as recognized by DP08.
      Taking then a value 
    for the stellar radius in the broad range $R=7-9 R_\odot$ , the equatorial 
    surface value,  $v_{eq}= 49\pm 3$ km/s and  
       the  observed  splitting of  $1.3032\mu$Hz yields a ratio 
   ${\Omega_{inner}/ \Omega_{surf}}$ in the range $[1.8-5]$  definitely
   indicating a non rigid rotation. There is not yet an agreement 
   concerning the radial order of the identified
   modes but the triplet seems in any case to be of mixed nature and therefore able to probe the
   core rotation. 
   DP08 did not consider overshoot and Desmet {\sl et al.} (2009) found that core overshoot must be smaller
   than 0.4 Hp.


\subsubsection{Summary}
These studies lead to the conclusion that a few rotationally split 
 modes can provide important  information about 
internal rotation and core overshoot of  $\beta$ Cephei stars  
if the modes are identified,  enough precise measurements are obtained and  
the age of the star  is  such that excited modes have mixed g, p  nature.
Trying to disentangle overshoot and rotation effects on core  element mixing
is only starting  with a measure of their relative magnitude  as is illustrated in Table 1.
As emphasized by DP08, in that respect, seismic modelling of fast rotators are needed.  
Once the size of the mixed core and the ratio of core to surface rotation are reliably determined, 
 the  next  issue is to estimate ,
 what part in the seismically measured extension of the core, $d_{ov}$, 
 comes from convective eddy overshooting and 
what part comes from other transport processes such as those induced by  rotation.



\subsection{ Splitting asymmetries  : distorsion}

\begin{figure}[t]
\centering
\includegraphics[height=6cm,width=10cm]{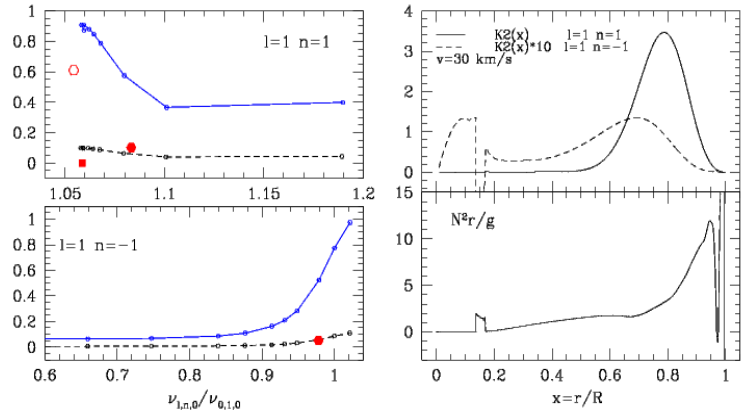}
\caption { {\bf left:} Scaled asymmetries ${\cal R}_m   ~10^{3}$
 for $\ell=1$ $n=1$ (top) and $n=-1$ (bottom) modes
  in function of the $m=0$ frequency  scaled by the radial fundamental mode frequency. 
  The  open dot (resp. full dot, full square) 
  represents the observed asymmetry for $\theta$ Oph, 
 (resp. for V386 Cen, $\nu$ Eri). The solid (resp. dashed)  line corresponds to 
 $v=30$ km/s (resp. 10 km/s)  $8.2 M_\odot$ models.  
 The central hydrogen content $X_c$ is  decreasing  toward the right.
{\bf top right:} Kernels $K_2(x)$ for splitting asymmetries  of $\ell=1, n=1$ (p1) mode (solid line)
  and  $\ell=1, n=-1$ (g1) mode (dashed line)  for model with $Xc=0.35$. 
The abscissae is the normalized
radius.{\bf bottom right} Run of the normalized Br\"unt-V\"aiss\"al\"a profile  $N^2r/g$ 
for the corresponding model with r/R.  {\it from Goupil \&  Talon, 2008)}}
\label{asym}
\end{figure}

The splitting asymmetry, $A_m$ (Eq.\ref{Am}), for acoustic modes 
  is mainly due to  the oblateness of the star caused by the centrifugal force although for low radial
  order modes, the Coriolis contribution is also significant. Fig.\ref{asym} represents the normalized splitting 
asymmetries: 
\begin{equation}
{\cal R}_m \equiv A_m/\nu_{0,1,0}
\end{equation}
for the $\ell=1, p_1$ and $g_1$ modes 
  in function of the  scaled frequency  $ y=  \nu_{\ell,n,0}/\nu_{0,1,0}$  
  where $\nu_{0,1,0}$ is the frequency of the radial fundamental mode.  
${\cal R}_m $ is plotted  for $\theta$ Ophiuchi, 
HD129929 and $\nu$ Eri.  The same quantities for 8.2 $M_\odot$ stellar models are also represented. 
The models have been
computed with CESAM2k code (Morel, 1997) assuming standard physics
 (Lebreton {\sl et al.}, 2008; Goupil 2008) 
including a core overshooting distance of 0.1  $H_p$  and an initial
hydrogen abundance $X=0.71$ and metal abundance $Z=0.014$. 
The evolution of the selected models is represented by the central hydrogen content $X_c$ from 0.5 to 0.2. 
The frequencies have been computed  using a second order perturbation method 
and an adiabatic  oscillation code WAR(saw)M(eudon) adapted
from the  Warsaw's oscillation code
(Daszy{\'n}ska-Daszkiewicz {\sl et al.}, 2002). For each model, two sets of frequencies are computed 
assuming  a uniform rotation 
corresponding  to $v=30$ km/s and $v=10$ km/s respectively. 
These sequences of models do not represent true evolutionary sequences 
as in realistic conditions, the rotation changes with time and  
can be non uniform.   They however  illustrate the evolution 
of the asymmetry when a mode
changes its nature during evolution, from pure p mode to mixed 
  p and g mode for instance.  
Indeed pure g modes have small asymmetries compared with 
pure p modes because they have much smaller
amplitude in the outer envelope where distorsion  has its most significant effect. This
is illustrated in Fig.\ref {asym}. In a perturbation description, 
  one finds that ${\cal R}_m $  is a second order effect  proportional to 
  $\Omega^2$ (DG92; Goupil {\sl et al.}, 2000; Goupil, 2009  and references therein).
The variation of   ${\cal R}_m $  with the scaled frequency $y$ 
(ie   with stellar  evolution)  is similar for the $v=30$
and $v=10$ km/s sequences of models  but 
${\cal R}_m $ is roughly 9 times (ie ratio of $ \Omega^2$) larger 
 for $v=30$ km/s models than $v=10$ km/s models.
 For pure p modes, the asymmetry   amounts to  
${\cal R}_m \sim 0.8 ~10^{-3}$  whereas for pure g modes it almost vanishes. 
${\cal R}_m$ for  the $\ell=1,n=1$  mode  decreases    for older models (larger y).  
The reverse happens for  the $\ell=1, n=-1$ mode. 
 The reason is that for young models, $\ell=1, n=1$ and
$n=-1$ modes are pure p and  g modes respectively. When the model is more evolved, these 2 modes 
 experience an avoided crossing  and exchange their nature. From a perturbative approach, one derives: 
\begin{equation}
A_m = \nu_{\ell,n,0} ~\int_0^1 \hat \Omega^2(x) ~K_2(x) ~dx
\end{equation}
where $\hat \Omega^2 = \Omega^2/(GM/R^3)$ and $x=r/R$ the radius normalized to the surface radius. $K_2(x)$ depends on 
the  centrifugal perturbation part of  pressure and density  as well as the
differential rotation $\Omega(x)$ and the mode eigenfunction.
Fig.\ref{asym}  shows $K_2(x)$ in function of the normalized radius $x=r/R$  
for  $\ell=1, n=1~ (p_1)$ and $\ell=1, n=-1 ~(g_1)$ modes for the $v=30$ km/s, 
8.2 $M_\odot$  model with $X_c=0.5$. 
The inner layers contribute to the asymmetry of  $\ell=1,g_1$ multiplet 
in contrast with the $\ell=1,p_1$ multiplet
for which the kernel  $K_2$ is concentrated toward the surface layers. The asymmetry of
the $\ell=1,g_1$  multiplet is sensitive to the  inner maximum of the Br\"unt-V\"aiss\"al\"a
frequency, 
arising from the $\mu$-gradient,  which contributes negatively to
$K_2$. 
As the negative contribution is very localized, it
decreases the asymmetry only slightly compared to a pure p mode for 
a uniform rotation. However, one
can expect a larger decrease in case of a 
rotation faster in the inner regions than the surface. \\

Theoretical estimates seem to disagree with observed asymmetries deduced from $\ell=2$ modes
  for $\theta$ Ophiuchi 
(Briquet {\sl et al.}, 2007)  and $\nu$ Eridani for $\ell=1, p_2$
 (Dziembowski \& Jerzykiewicz, 1999,  PHD04, Suarez {\sl et al.} 2009).
Is the disagreement real? The question has some relevance as the asymmetry values
 are only marginally above the observation uncertainties. 
Or can it be that 
 the observed frequencies do not belong to the same multiplet  as suggested by DP08
for $\nu$ Eri?

\subsection{ Axisymmetric modes: mixing}

 Rotationally induced mixing of chemical elements changes the structure and in  
 particular affects the Br\"unt-V\"aiss\"al\"a
frequency $N$ at the border of the convective core.
  As a consequence, at a given location in a HR diagram corresponding to  an
  observed star, one can find several models with different structures and
  therefore likely different values of the  mode   frequencies including
  axisymmetric modes which can then be used as diagnostics for mixing.


\begin{figure}[t]
\epsfig{file=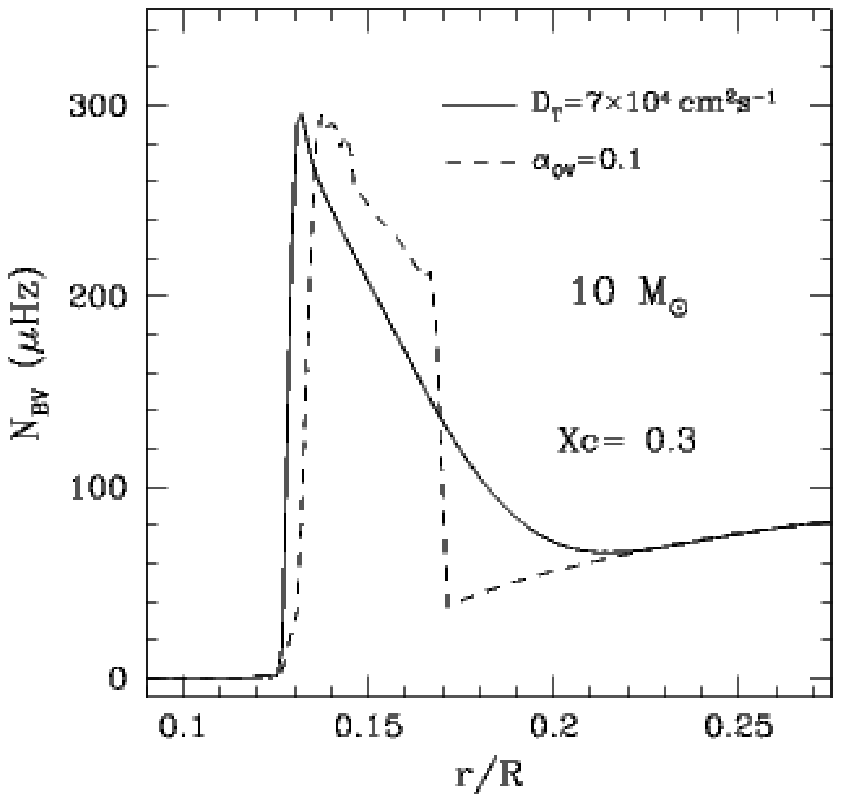,width=6cm} 
\epsfig{file=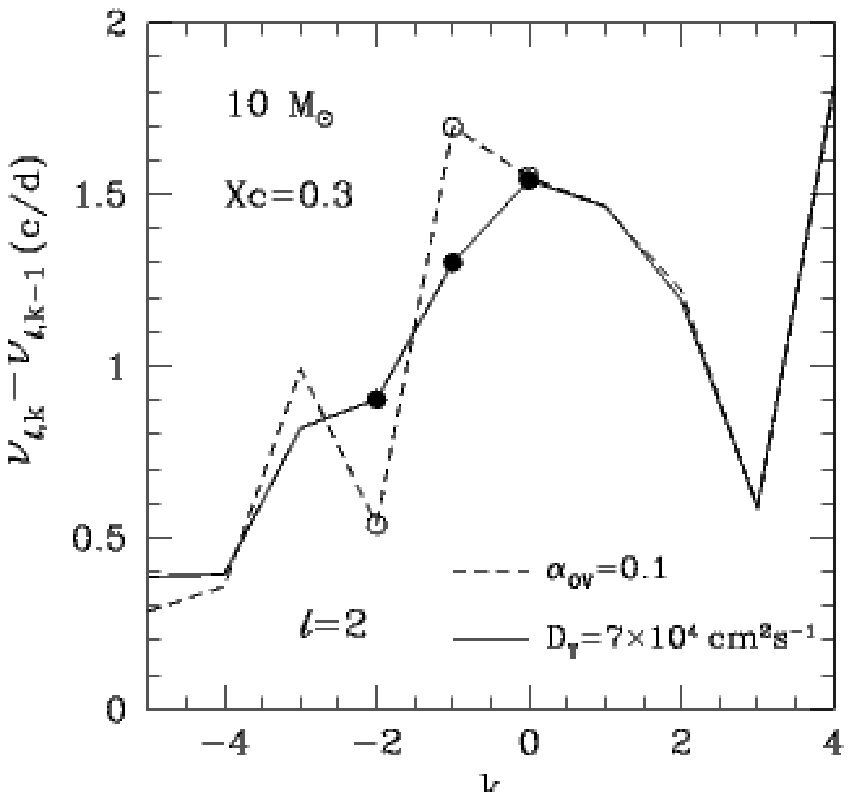,width=6cm} 
\caption {{\bf left:} Br\"unt-V\"aiss\"al\"a profile in the central region of a 10 $M_\odot$ model with 
$X_c=0.3$  and an initial velocity of 50 km/s. 
{\bf right:} the large separation $\nu_{\ell,n,0}-\nu_{\ell,n-1,0}$
 in function of the radial order  n  for $\ell=2$ modes for a model including 
turbulent  mixing (solid) line and a model including a $0.1 ~H_p$ overshoot instead (dashed line). {\it (from Montalban et al. 2008)}}
\label{mj16}
\end{figure}

\subsubsection{Uniform and constant  diffusion coefficient $D_t$:}
Montalban {\sl et al.} (2008) and Miglio {\sl et al.} (2008) 
 investigated the effect of turbulent
mixing on a g-mode frequency spectrum 
and the ability of such modes to probe the size of 
stellar convective cores. 
They assumed 
 a constant in time and uniform in space    
global diffusion coefficient  $D_t= D_{eff}+D_v$ in Eq. \ref{cj} above. 
The constant value for $D_t$ is chosen so as to correspond 
to the value near the convective core provided by  a  Geneva stellar 
model including rotationally induced mixing.
This is valid for g-modes which have most of their amplitude
there (see Miglio  {\sl et al.},  2008) 
 The model is a mid main sequence ($X_c=0.3$)  10 $M_\odot$
 with $D_t =7~10^4$\,cm$^2$/s chosen to correspond 
 to a rotational velocity  $v $= 50 km/s.

Fig.\ref{mj16} shows the
 Br\"unt-V\"aiss\"al\"a frequency ($N$) profile for a model with turbulent chemical 
 element mixing and a  model with no turbulent chemical 
 element mixing but  including instead   core overshoot assuming an overshoot distance of 0.1 $H_p$.
Differences can be seen at the edge of the convective core. The  Br\"unt-V\"aiss\"al\"a  frequency
of the model  with  turbulent  mixing behaves more smoothly in the $\mu$-gradient
 region above the convective core than  for the model 
computed  with no turbulent mixing but with an  overshoot distance  of 
$0.1\,H_p$. From Geneva code calculations,  the evolution of the 
rotation profile leads to  a core to envelope ratio of 1.6.
The differences between the two profiles arising at the edge of the
convective core  cause significant  changes  on frequencies  of g-modes and mixed modes. The  frequency separations 
$\Delta_{n,\ell}=\nu_{\ell,n,0}-\nu_{\ell,n-1,0} $  differ by    a few $\mu$Hz
  for radial order $n =-1$ and $n=-2$ , $\ell=2$ modes
 between the model with overshoot  $0.1\,H_p$  and the model with turbulent mixing
(Fig.\ref{mj16}). At higher frequencies for pure p-modes,  
no differences in $\Delta_{n,\ell}$ are seen when adding  turbulent
 mixing or not.


\begin{figure}[t]
\centering
 \epsfig{file=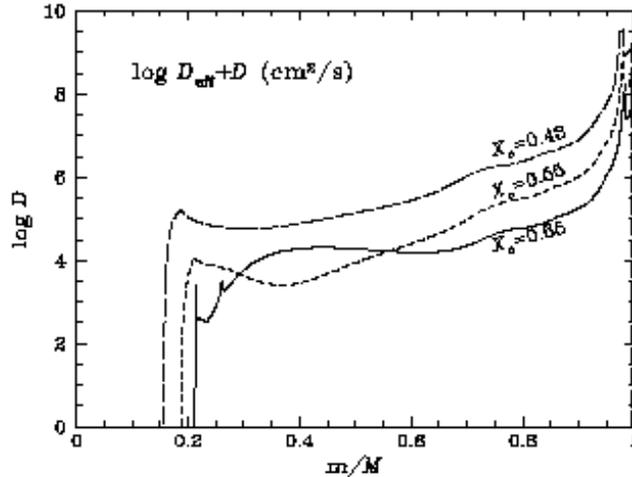,width=9cm}
 {\vskip -.5truecm}
\caption { Run of the rotationally induced  
  turbulent  coefficient, $D_t$,  with the relative shell  mass at 3 different
  evolutionary stages with  ages 0.5 Myr, 1 Myr and 1.5 Myr respectively
and    labelled with their central hydrogen content $X_c$-  
  leading to  the stellar model  {\bf $V_{15}$} ($X_c=0.3$) {\it (from Goupil \& Talon 2008)}.
  }
\label{Deff}
\end{figure}

 \subsubsection{Rotationally induced diffusion coefficient }\label{rotinduced}

In this section, we consider  stellar models which are computed 
with the Toulouse-Geneva evolutionary code
which  includes  the coupling between rotationally induced mixing and momentum
transport (Eq.\ref{cj} and Eq.2 above)  as described by Talon (2008).
The rotational evolution of the star begins from solid
 body when the core is still radiative, 
 shortly after the star leaves the Hayashi track. 
 A $8.5\,M_\odot$ 
mass has been chosen so that  the models  evolve through 
 the HR   diagram to a location  where the star 
$\theta$ Ophiuchi  is expected
($\log L/L_\odot = 3.73$, $T_{\rm eff} =   4.35$). This corresponds to a mid main
sequence model,  {\bf $V_{15}$}, with a central hydrogen content  $X_c =0.3$. 
The evolution has been initiated  
with a   uniform rotational velocity  $v=15$ km/s on the pms; 
the rotation profile then evolves to strongly
differential rotation  so that
 {\bf $V_{15}$} has  a   surface velocity of $v=48.2$ km/s   and a ratio
 $\Omega_{core}/\Omega_{surf}=1.6$
 when crossing the $\theta$ Ophiuchi  location in the HR diagram at  an age of 
 19.65 Myr.

\begin{figure}[t]
\centering
\epsfig{file=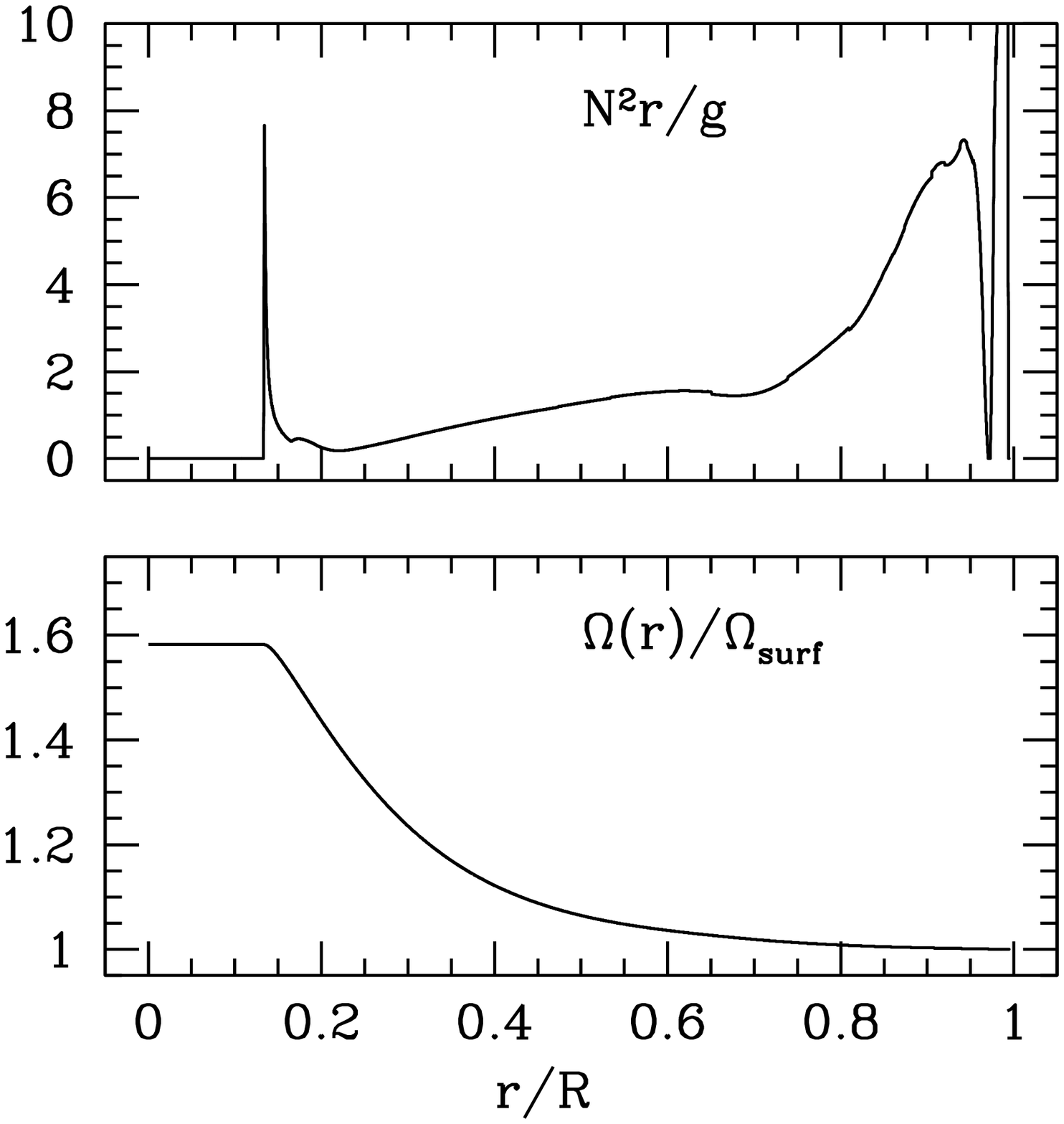,width=8cm} 
\caption {Model {\bf $V_{15}$} with a surface rotational velocity v=48.2 km/s. 
{\bf top:}  Profile of  of the normalized Br\"unt-V\"aiss\"al\"a  frequency 
as defined by $N^2 ~r/g$ in function of the normalized radius $r/R$.   
{\bf Bottom:} rotation
profile  normalized to its surface value. The core to surface ratio for the  rotation
rate then  is 1.6  {\it (from Goupil \& Talon 2009 in prep.)}.  
}
\label{rot}
\end{figure}

 The  diffusion coefficient, $D_t$, depends on the
meridional circulation velocity and the local turbulence strength. It   varies with
depth  and evolves with time as illustrated in Fig.\ref{Deff}. 
The $D_t$ profile is represented for  3 models   with ages 0.5 Myr, 1 Myr and 1.5 Myr  built assuming 
an initial 15 km/s velocity on the pms.  The rotation evolving from uniform to strongly 
differential rotation  causes a relaxation toward a
stationary profile which persists with only an  ajustement due
to expansion and contraction with evolution (Goupil \& Talon 2008). \\
Effect of rotationally  induced  mixing on the structure  is significant at the edge of the
  convective core as emphasized in Fig.\ref{Deff} where we compare the  squared   
  Br\"unt-V\"aiss\"al\"a  profile, $N^2$,
 in the vicinity of the  edge of convective core for model {\bf $V_{15}$} 
  and  a model {\bf $ V_0$} which includes neither  
rotationally induced mixing nor  overshoot. Inclusion of  rotationally  induced mixing leads to the model 
 ${\bf V_{15}}$
which shows a narrower maximum of  Br\"unt-V\"aiss\"al\"a 
  profile at the edge of the convective core compared with
that of  ${\bf V_0}$. 

\begin{figure}[t] 
 \epsfig{file=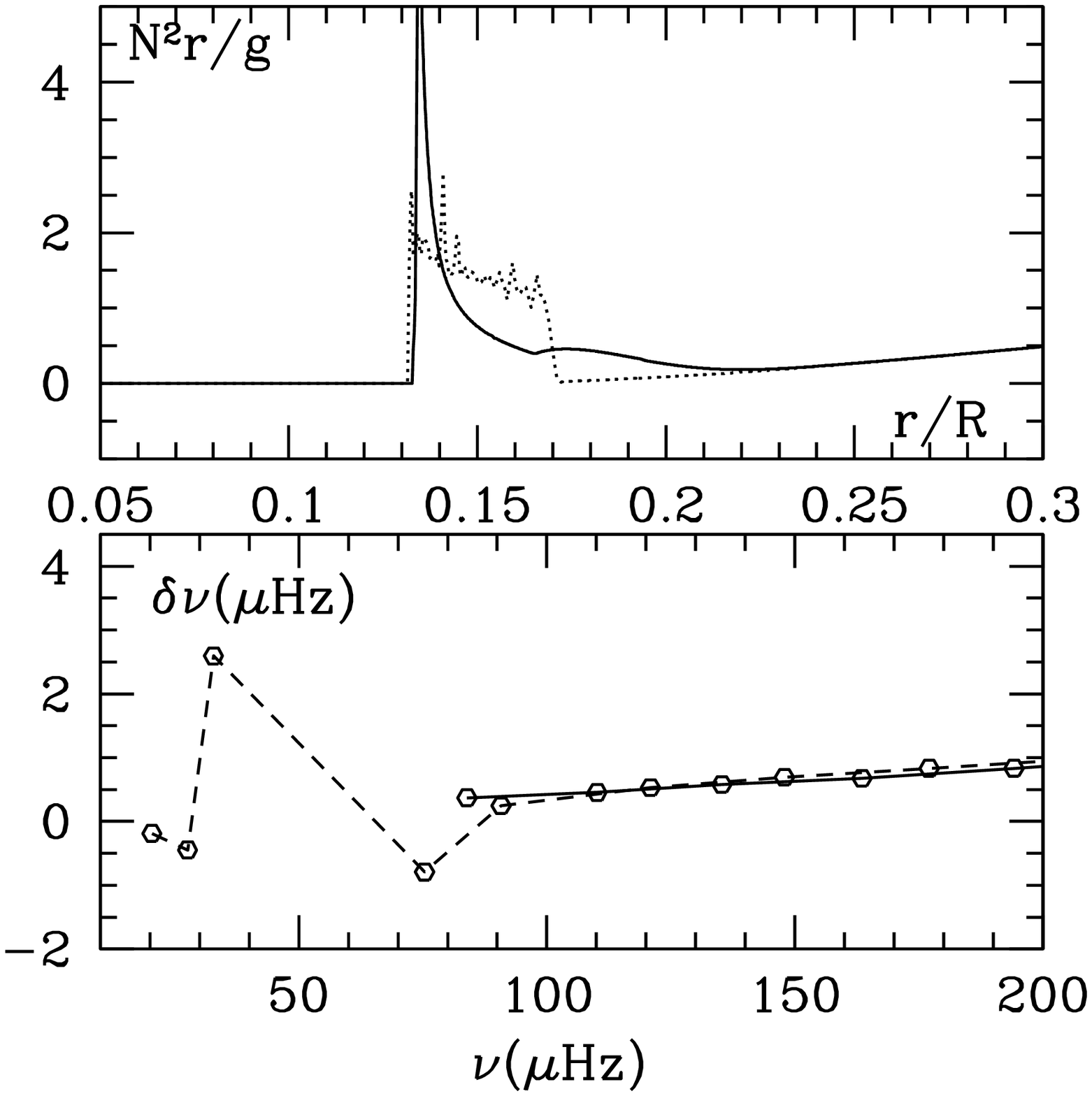,width=6cm}
\epsfig{file=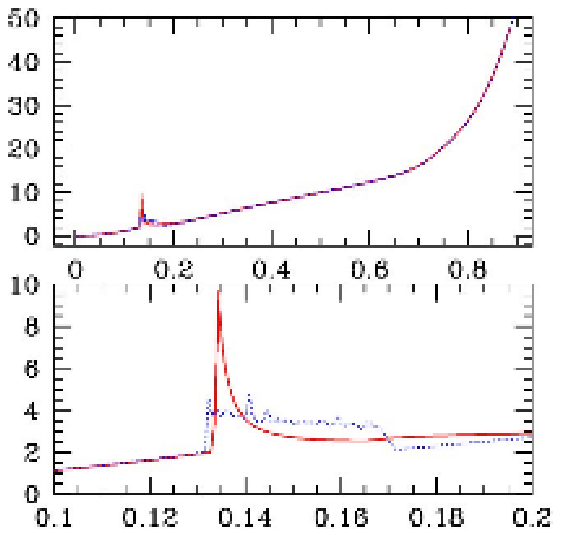,width=6cm} 
 \caption { {\bf left top:}  Zoom of Br\"unt-V\"aiss\"al\"a frequency  profile 
 in the vicinity of the  edge of convective core in function of the normalized radius $r/R$
  for model  {\bf $V_{15}$} (dashed line) and model  {\bf $V_0$} (solid line).
The local 
 maximum of $N^2r/g$ corresponds to a nonzero $\mu$-gradient. It decreases more sharply in
 presence of rotationally induced mixing because mixing results in smoothing the $\mu$-gradient.
  {\bf Left bottom: } Differences $\delta \nu = \delta \omega/(2\pi) $ 
 between frequencies computed from model {\bf $V_0$} (no rotationally
  induced mixing included) and model 
  {\bf $V_{15}$} for $\ell=0$ (solid line) and $\ell=1$ (dashed line)
 $m=0$ modes in $\mu$Hz {\sl (from Goupil \& Talon, 2008)}.
{\bf right:} $ d\ln \rho/d\ln r$ in function of the normalized radius $r$  for model
 {\bf $V_{15}$} (dashed line)and model {\bf $V_0$} (solid line)
{\bf right top:} from center to surface,  {\bf right bottom:} in the central region 
{\sl (from Goupil \& Talon 2009, in prep.)}
}
\label{mj21}
\end{figure}
To illustrate the impact of such a difference on the oscillation frequencies,
 we  compare low radial order frequencies  of the  models ${\bf V_{15}}$,  and ${\bf V_0}$. 
Modes $p_1$, $p_2$, $g_1$ for these models have  amplitudes near the edge of the convective core.
Fig.\ref{mj21} shows that  this can result in  significant frequency differences for the same mode 
easily
detectable with CoRoT observations. The frequencies of these
modes are quite sensitive to the detail of the Br\"unt-V\"aiss\"al\"a profile in this region.
This means  that some care must be taken when computing these frequencies and drawing
 conclusions. The
frequencies of these modes   are indeed sensitive  not only to the
physics but unfortunately also to the numerics which can be quite inaccurate in this 
region of the star.

The sign and magnitude of $\delta \omega= \omega_{V15}-\omega_{V0}$  are dependent on the mode 
when it has amplitude in the regions where the nonrotating model
and the model with rotationally induced mixing differ.  
We consider here, as in Sec.3.5, only the effect of rotationnally
 induced mixing on the spherically symmetric structure. Differences  in the structure
  of the  model  {\bf $V_{15}$}   which includes rotationnaly induced mixing 
and the  model  {\bf $V_0$} which does
not  result in differences in the eigenfrequencies
which we note $\delta \omega = \omega_{0,\Omega}-\omega_{0,\Omega=0}$.

The structure of the  models {\bf $V_{15}$} and {\bf $V_0$} indicates that $p_\Omega$  and its derivative,
 the gravity $g_\Omega$, the density $\rho_\Omega$  
are  not significantly modified compared to the derivative of the density. 
Fig.\ref{mj21}  shows that the largest difference 
 $\delta({d\log \rho_\Omega/ d\log r}) $ arises near the convective core.
Accordingly from  Eq.\ref{eqcent5}, one expect larger 
frequency differences  $\delta \omega$  for  mixed modes compared to p-modes. 
 This is what is observed in Fig.\ref{mj21}.
As explained in Sect.3.5, with the help of the integral relation for $\delta \omega$,
  the frequency differences
 for  high frequency (i.e.  pure) p-mode   is small and positive. 
 For lower frequency mixed modes, 
 $\delta   \Bigl({d\ln \rho_\Omega \over d\ln r}\Bigr)$  can be positive  and the  frequency
 difference can be large and  negative as illustrated in Fig.\ref{mj21}.

\section{ Cubic order versus latitudinal dependence}

It has been known for a long time that latitudinal variations of the rotation
rate generate departures from linear splitting. On the other hand,
a fast uniform rotation can generate cubic order  corrections  to 
the frequency of non axisymmetric modes which
also cause departure from linear splitting.
The latitudinal  correction to the linear splitting 
is proportional to   the $\Omega$ gradient whereas cubic order effects, as their
name indicate,  are proportional to $\Omega^3$. 
It is expected that the dependence of these corrections  with the frequency 
differs when it is due to latitudinal differential rotation or to cubic effects.

Low mass stars are known to be slow rotators. Indeed  due to their outer convection
zone,  they undergo   magnetic 
braking. Due again to their outer convection zone, obervational evidences exist for  
surface latitudinal differential rotation. Hence for these stars, the averaged rotation rate  $\Omega$ 
is small and $ \Delta \Omega= \Omega_{equa}-\Omega_{pole}$, the difference between the rotation rates at the
equator and the poles, can be large 
(25\% -30\% for the Sun, between 1\% and 45\% for a star like  Procyon, Bonanno {\sl et al.} 2007). One therefore expects that latitudinal corrections to the splittings 
dominate over cubic order ones which are negligible. 
On the other hand, more massive stars on the main sequence have shallower
convection zones  which  even disappear above $\sim $ 3- 5 $M_\odot$.  These stars 
usually are fast rotators  with a radiative envelope  which may or may not be
in latitudinal differential rotation.  For these fast rotators, one can wonder
what is the  minimal latitudinal shear which  dominates over 
cubic order effects and can therefore be detectable.
Here we quantify this issue with the help of a polytropic model with index 3. The
constants characterizing the polytrope are   taken  to correspond to   model {\bf A} considered  
 in Sect.\ref{rotinduced}.  
We  establish first  the splitting correction due to latitudinal differential
rotation. This is then compared with the  splitting correction arising from
cubic order effects as derived by previous works. We assume a rotation velocity of 100
km/s.

\subsection{Latitudinal dependence}


Hansen {\sl et al.} (1977) derived the expression for  the
rotational  splitting of adiabatic nonradial
 oscillations for slow  differential (steady, axially symmetric) rotation  $\Omega(r,\theta)$ 
 and applied it  to numerical models of white dwarfs and 
 of massive main sequence stars   assuming a cylindrically symmetric rotation law.
In the solar case, the  effects of latitudinal differential
 rotation on theoretical frequencies  were investigated  by Gough \& Thompson (1990), 
Dziembowski \& Goode (1991, DG91) and  DG92 who also considered the case of 
$\delta$ Scuti stars.

In order to be able to compute the splittings from Eq.\ref{Sm} and Eq.\ref{sp}, one must specify a rotation
law. It is convenient to assume  a rotation of the type:
\begin{equation}
\Omega(r,\theta)=  \sum_{s=0}^{s_{max}} ~ \Omega_{2s}(r) ~(\cos\theta)^{2s}
\label{latit1}
\end{equation}
where $\theta$ is the colatitude and we take $s_{max}=2$. 
The surface rotation at the equator is $\Omega(r=R,\theta=\pi/2) =\Omega_0(r=R)$. 
 Note that in the solar case, $\Omega_2$, $\Omega_4$ are 
 negative and the equator rotates faster
than the poles  (DG91, Schou \etal,  1998). As shown in Appendix, 
inserting Eq.\ref{latit1} into Eq.\ref{sp} yields 
the following  expression for the generalized splitting
 (Eq.\ref{sm84} in Appendix): 
 \begin{eqnarray}
S_m  &=&  \int_0^R   \Omega_0(r) ~K(r) ~dr+ \sum_{s=0}^{s=2}  m^{2s} ~ H_{s}(\Omega) 
 \end{eqnarray}
with $K(r)$ defined in Eq.\ref{Kr} and 
  \begin{eqnarray}
& &H_{s} (\Omega) =\nonumber  \\ -
& & {1 \over I}  \int_0^R  ~\Omega_0(r)~ \Bigl[ R_s ~\Bigl(\xi^2_r  -2\xi_r \xi_h+ \xi^2_h   (\Lambda-1)\Bigr) +Q_s ~ \xi^2_h\Bigr)   \Bigr]  \rho_0 r^2
dr 
 \end{eqnarray}
where $R_s$ and $Q_s$ depend on $\Omega_2,\Omega_4$ 
and $\Lambda=\ell(\ell+1)$ and are given by Eq.\ref{SS81} and Eq.\ref{SS82} (Appendix) respectively.

\medskip

\ni {\bf  \small a) Uniform rotation} In that case, 
  $\Omega(r,\theta)= \Omega_0, ~ \forall r$, $\theta$; $\Omega_2,\Omega_4=0$
i.e.  $R_j, Q_j =0$ for $j=0,2$ hence $H_{m,j}=0$. One recovers the well known expression:
 \begin{eqnarray}
S_m  &=&       \Omega_0 ~ \beta  
  \end{eqnarray}
  where, for later purpose,  we have defined 
 \begin{eqnarray}
 \beta &=&      \int_0^R ~  K(r)~ dr  \nonumber  \\
  &=&   -{1  \over I}  \int_0^R ~  \Bigl[  \Bigr.
 \xi_r^2   - 2\xi_r \xi_h     + (\Lambda-1)~ \xi_h^2 \Bigl.  \Bigr] ~\rho_0 r^2 dr
\label{beta}
  \end{eqnarray}
This is usually rewritten as :
$$S_m =  \Omega_0 ~(C_L-1)   $$
where $C_L$ is the Ledoux (1951) constant
$$ C_L  =  { 1 \over I}    \int_0^R ~ \Bigl[ 2\xi_r \xi_h +  \xi_h^2  \Bigr] ~\rho_0 r^2 dr  \nonumber
$$
\medskip

{\bf \small b) Shellular rotation}
then $\Omega(r,\theta)= \Omega_0(r)$ and $s_{max}=0$; again here: $\Omega_2,\Omega_4=0$
ie  $R_j= Q_j=0$  for $j=0,2$ and 
 \begin{eqnarray}
S_m  &=& - {1  \over I}      \int_0^R ~\Omega_{0}(r)~  \Bigl[  \Bigr.
 \xi_r^2   - 2\xi_r \xi_h     + \Lambda ~ \xi_h^2 \Bigl.  \Bigr] ~\rho_0 r^2 dr  
  \end{eqnarray}

\medskip

{\bf \small c) Latitudinally differential rotation only}
In that case, $\Omega_{2j}, j=0,2$ are depth independent and  $R_s$ and $Q_s$ are constant  and 
  \begin{eqnarray}
S_m  =   \Omega_0 ~ \beta  +  \Omega_0 ~\sum_{s=0}^{s=2}  m^{2s} ~  (R_{s}(\Omega) ~\beta + Q_{s}(\Omega)~ \gamma )
 \end{eqnarray}
with $\beta$ defined in Eq.\ref{beta}
and 
$$ \gamma= - {1 \over I}\int_0^R ~ \xi^2_h ~ \rho_0 r^2 ~dr  $$

For a   triplet $\ell=1$, $m=1$ ($\Lambda=2$)  then
 \begin{eqnarray}
S_1  &=&   \Omega_0 ~ \beta  +  \Omega_0 ~  (R(\Omega) ~\beta + Q(\Omega) ~ \gamma) 
\label{S111}
 \end{eqnarray}
with (using Eq.\ref{SS81} and  Eq.\ref{SS82}):
 \begin{eqnarray}
 R(\Omega) &=& \sum_{s=0}^{s=2}   ~ R_{s}(\Omega)=  {1\over 5}{\Omega_2\over \Omega_{0}} + {3\over 7}~  
 {\Omega_{4}\over\Omega_{0}}  \\
Q(\Omega) &=& \sum_{s=0}^{s=2}   ~ Q_{s}(\Omega)=   - {24\over 5} {\Omega_{4}\over\Omega_{0}}  
  \end{eqnarray}

In the solar case, $\beta \sim -1$ and  $|\beta|  >>|\gamma|$  for the excited high frequency  p-modes.  
 \begin{eqnarray}
S_1  &\approx &  - \Omega_0 ~ \Bigl(1+ {1\over 5}{\Omega_2\over \Omega_{0}} + {3\over 7}~ 
  {\Omega_{4}\over\Omega_{0}}  \Bigr)  
\label{S112}
 \end{eqnarray}
With $\Omega_2/\Omega_0=-0.127, \Omega_4/\Omega_0=-0.159$ (from DG89), one obtains a  departure 
from linear splitting of   
 $|{S_1 / \Omega_0} +1|=0.093 $ i.e. a $\approx$ 10\% change in the solar case.
 For upper main sequence stars, excited modes are around the fundamental radial mode
 and may be mixed modes with $|\beta| \sim |\gamma| \sim 1/2$. This leads for instance to 
 $|{S_1 / \Omega_0} +1/2|\approx 5 \% $  for $\Omega_2/\Omega_0 $ and $\Omega_4/\Omega_0$ equal 
 to 1/5 of the solar
 values.
 
 \subsection{Latitudinal dependence versus cubic order effects  }
 
Let assume on one side a pulsating star uniformaly rotating
 with a rate $\Omega_0$ high enough  that cubic order ($O(\Omega_0^3)$) contributions 
are significant. On the other side, one also considers    a model rotating with 
a latitudinally differential  rotation (uniform in radius).
One issue then is which  one of these 
two effects dominate over the other one
since the cubic one is $O(\Omega^3)$ whereas the other one is $O(\Delta \Omega)$ ?
For stars other than the Sun, one can simply  assume the rotational latitudinal shear $\Delta \Omega=\Omega_2$  
with $\Omega_4=0$ and  $\Omega(\theta)= \Omega_0+ \Delta \Omega  \cos^2\theta $.
For $\ell=1$ modes, Eq.\ref{S111}  becomes 
\begin{equation}
S_1 (lat)=  \Omega_0 ~ \beta \Bigl( 1+  {1\over 5 } {\Delta \Omega \over \Omega_0} \Bigr)
\label{S1lat}
\end{equation}

Expressions for the frequency correction (in rad/s)  for cubic order effects assuming a uniform rotation 
has been derived by Soufi {\sl et al.} (1998). 
Part of the cubic order effet  is included in the eigenfrequency $\omega_{0,\Omega}$ and therefore  
is also included in second order coefficients which indeed involve $\omega_{0,\Omega}$. Another part of the 
cubic order effects is  included  as an additive correction to the frequency.

Frequency  up to 3rd order were computed for models of $\delta$ Scuti stars by 
Goupil {\sl et al.} (2001),  Goupil \& Talon (2002), Pamyatnykh (2003), Goupil {\sl et al.} (2004).
Karami (2008) rederived the cubic order effects following  Soufi et al.'s  approach 
and Karami (2008, 2009) applied it to
a ZAMS  model of a 12 $M_\odot$ $\beta$ Cephei star. 
He found  that cubic order effects are of the order of 
$0.01\%  $ for a $l=2,n=2$ and $0.5\%$  for a n=14 mode  for a 100 km/s
 rotational velocity. 
Values of the third order additive correction to the frequency 
  were listed  for  $\ell=1$ p-modes of a polytrope of index 3 by  Goupil (2009).

Here we write the  splitting  under the 
form:
\begin{equation}
S_m (cubic) =  \Omega_0 ~ \beta+ \Omega_0 ~\Bigl({\hat \Omega_0\over \sigma_0}\Bigr)^2 ~T_{|m|}
\end{equation}
where 
the last term represents   the {\it full} cubic order  
contribution with  $\sigma_0$ is the normalized frequency 
 of the nonrotating polytrope and  $\hat \Omega_0 = \Omega_0 /\Omega_K$.

Tab.\ref{tab1} lists the value of the dimensionless  coefficients
  $T_1/\sigma^2_0$  and $-\beta, -\gamma$ for $\ell=1$ modes for a polytrope with a polytropic index 3. 
 The  coefficient ${T_1 / \sigma^2_0}$ remains 
nearly constant with increasing frequency for 
frequencies above $\sigma_0 > 10$ i.e. for p modes
For $\sigma_0 > 10$ (p-modes), $-\beta\approx 1$ 
and   $T_1/\sigma^2_0 \approx -0.09$. The splitting is decreased by a
latitudinal dependence with $\Delta \Omega<0$ whereas it is increased by cubic
order effects $T_1/\beta>0$ .
In absolute values, the effect of latitudinal differential rotation on the splittings then 
dominates over cubic order effects 
whenever:
$$|{\Delta \Omega \over \Omega_0}|  > 0.45~  \hat \Omega^2_0  $$
For model {\bf A} and a rotational velocity 100 km/s, $\hat \Omega_0 =0.174$ then 
$ |{\Delta \Omega \over \Omega_0}|  > 1.36\%$
For a faster rotator with for instance 200 km/s,the latitutinal shear must be larger 
i.e. $ |{\Delta \Omega \over \Omega_0}|  >~5.45 \%$.



For the slowly rotating 
$\beta$ Cep stars considered in Sect.4  above ($v <50$ km/s), 
cubic order effects  in  the splittings  
can  be  neglected in front of  latitudinal effects  equal or larger than 
$0.34 \%$. At this low level, both effects are comparable to  the observational
uncertainties ($0.1\%$).  

\begin{table}[]
\caption{ Coefficients   assuming a uniform rotation
 for a polytrope with polytropic index $3$ and  adiabatic index $\gamma=5/3$. The
squared frequency $\sigma^2_0$ is the dimensionless squared frequency $\omega^2/(GM/R^3)$.  
Spherical centrifugal distorsion of the polytrope has not been included.}
\label{tab1}
\begin{center}
\begin{tabular}{llllllllll}  
\hline
      &                   &                &          &  $\ell=1$   &  &     \cr
\hline
   n &   \,\,  $\sigma^2_0$ & \,\, $C_{L0}$ &\,\, $X_1$  & \,\,$X_2$ &\,\, $Y_1$  & \,\, $Y_2$  & \,\,
   $T_1/\sigma^2_0 $ &  \,\, -$\beta$ &   \,\, -$\gamma$    \cr
\hline
  -7 &  \,\,    0.22 &  \,\,   0.479 &  \,\,   0.417 &  \,\,   0.008 & \,\,    0.012&\,\,   -0.018 &  \,\,
    0.592 &  \,\,    0.521   &  \,\,  0.456  \\
  -6 &  \,\,    0.28 &  \,\,   0.476 &  \,\,   0.419 &  \,\,   0.005 & \,\,    0.015&\,\,   -0.023 &  \,\,
    0.462 &  \,\,    0.524    &  \,\, 0.450   \\
  -5 &  \,\,    0.37 &  \,\,   0.473 &  \,\,   0.422 &  \,\,   0.001 & \,\,    0.020&\,\,   -0.029 &  \,\,
    0.351  &  \,\,    0.527  &  \,\,   0.441   \\
  -4 &  \,\,    0.52 &  \,\,   0.469 &  \,\,   0.425 &  \,\,  -0.004 & \,\,    0.026&\,\,   -0.039 &  \,\,
   0.254   &  \,\,   0.531  &  \,\,   0.431  \\
  -3 &  \,\,    0.78 &  \,\,   0.466 &  \,\,   0.427 &  \,\,  -0.013 & \,\,    0.038&\,\,   -0.056 &  \,\,
   0.164  &  \,\,    0.534  &  \,\,   0.410\\
  -2 &  \,\,    1.28 &  \,\,   0.466 &  \,\,   0.428 &  \,\,  -0.024 & \,\,    0.059&\,\,   -0.089 &  \,\,
   0.073  &  \,\,    0.535   &  \,\,  0.386 \\
  -1 &  \,\,    2.51 &  \,\,   0.473 &  \,\,   0.422 &  \,\,  -0.035 & \,\,    0.106&\,\,   -0.159 &  \,\,
   -0.025 &  \,\,    0.528   &  \,\,  0.269 \\
   1 &  \,\,   11.37 &  \,\,   0.029 &  \,\,   0.777 &  \,\,   0.877 & \,\,    2.890&\,\,   -4.335 &  \,\, 
    0.024   &  \,\,    0.970  &  \,\,   0.025 \\
   2 &  \,\,   21.49 &  \,\,   0.034 &  \,\,   0.773 &  \,\,   0.864 & \,\,    5.802&\,\,   -8.703 &  \,\,
    -0.034  &  \,\,   0.966  &  \,\,   0.028\\
   3 &  \,\,   34.83 &  \,\,   0.033 &  \,\,   0.773 &  \,\,   0.851 & \,\,    9.624&\,\,  -14.436 &  \,\,
    -0.063  &  \,\,  0.966   &  \,\,  0.027  \\
   4 &  \,\,   51.39 &  \,\,   0.031 &  \,\,   0.776 &  \,\,   0.840 & \,\,   14.340&\,\,  -21.511 &  \,\,
    -0.077  &  \,\,    0.969  &  \,\,   0.026 \\
   5 &  \,\,   71.15 &  \,\,   0.027 &  \,\,   0.778 &  \,\,   0.832 & \,\,   19.940&\,\,  -29.909 &  \,\,
    -0.084  &  \,\,    0.973  &  \,\,   0.025 \\
   6 &  \,\,   94.09 &  \,\,   0.024 &  \,\,   0.781 &  \,\,   0.826 & \,\,   26.414&\,\,  -39.621 &  \,\,
   -0.088  &  \,\,    0.976  &  \,\,   0.023 \\
   7 &  \,\,  120.19 &  \,\,   0.021 &  \,\,   0.783 &  \,\,   0.821 & \,\,   33.757&\,\,  -50.635 &  \,\,
   -0.089 &  \,\,   0.979   &  \,\,  0.022 \\
   8 &  \,\,  149.43 &  \,\,   0.019 &  \,\,   0.785 &  \,\,   0.817 & \,\,   41.964&\,\,  -62.946 &  \,\,
   -0.089  &  \,\,   0.981  &  \,\,   0.020 \\
   9 &  \,\,  181.81 &  \,\,   0.017 &  \,\,   0.787 &  \,\,   0.814 & \,\,   51.032&\,\,  -76.548 &  \,\,
   -0.089 &  \,\,    0.984  &  \,\,   0.019 \\
  10 &  \,\,  217.32 &  \,\,   0.015 &  \,\,   0.788 &  \,\,   0.811 & \,\,   60.958&\,\,  -91.437 &  \,\,
  -0.089  &  \,\,    0.985  &  \,\,   0.018 \\
  11 &  \,\,  255.94 &  \,\,   0.013 &  \,\,   0.789 &  \,\,   0.809 & \,\,   71.739&\,\, -107.609 &  \,\,
  -0.088  &  \,\,   0.987   &  \,\,  0.017 \\
  12 &  \,\,  297.67 &  \,\,   0.012 &  \,\,   0.790 &  \,\,   0.807 & \,\,   83.375&\,\, -125.062 &  \,\,
  -0.087  &  \,\,  0.988  &  \,\,   0.017\\
  13 &  \,\,  342.51 &  \,\,   0.011 &  \,\,   0.791 &  \,\,   0.805 & \,\,   95.862&\,\, -143.793 &  \,\,
  -0.087    &  \,\,   0.989   &  \,\,  0.016\\
  14 &  \,\,  390.44 &  \,\,   0.010 &  \,\,   0.792 &  \,\,   0.804 & \,\,  109.201&\,\, -163.802 &  \,\,
  -0.086   &  \,\,   0.990   &  \,\,  0.015  \\
  15 &  \,\,  441.47 &  \,\,   0.009 &  \,\,   0.793 &  \,\,   0.803 & \,\,  123.392&\,\, -185.087 &  \,\,
  -0.085   &  \,\,   0.991  &  \,\,   0.014\\
  16 &  \,\,  495.59 &  \,\,   0.008 &  \,\,   0.793 &  \,\,   0.802 & \,\,  138.432&\,\, -207.648 &  \,\,
   -0.085  &  \,\,   0.992  &  \,\,   0.014 \\
  17 &  \,\,  552.80 &  \,\,   0.008 &  \,\,   0.794 &  \,\,   0.801 & \,\,  154.323&\,\, -231.484 &  \,\,
  -0.084   &  \,\,   0.993  &  \,\,   0.013 \\
  18 &  \,\,  613.09 &  \,\,   0.007 &  \,\,   0.794 &  \,\,   0.800 & \,\,  171.064&\,\, -256.595 &  \,\,
  -0.084  &  \,\,    0.993  &  \,\,   0.013\\
  19 &  \,\,  676.47 &  \,\,   0.006 &  \,\,   0.795 &  \,\,   0.799 & \,\,  188.655&\,\, -282.982 &  \,\,
  -0.083  &  \,\,   0.994  &  \,\,   0.012 \\
  20 &  \,\,  742.93 &  \,\,   0.006 &  \,\,   0.795 &  \,\,   0.798 & \,\,  207.097&\,\, -310.645 &  \,\,
  -0.083  &  \,\,     0.994  &  \,\,   0.012 \\
  21 &  \,\,  812.46 &  \,\,   0.006 &  \,\,   0.796 &  \,\,   0.798 & \,\,  226.389&\,\, -339.584 &  \,\,
 -0.082   &  \,\,   0.995   &  \,\,  0.012 \\
  22 &  \,\,  885.08 &  \,\,   0.005 &  \,\,   0.796 &  \,\,   0.797 & \,\,  246.534&\,\, -369.800 &  \,\,
  -0.082   &  \,\,   0.995  &  \,\,   0.011 \\
  23 &  \,\,  960.78 &  \,\,   0.005 &  \,\,   0.796 &  \,\,   0.797 & \,\,  267.530&\,\, -401.295 &  \,\,
 -0.082  &  \,\,   0.995  &  \,\,   0.011\\
\hline
\end{tabular}
\end{center}
\end{table}

\bigskip


\section{Conclusions}

We have seen along this review that several efficient
seismic tools  can be designed   to obtain valuable information on 
 the internal structure and
dynamics of  main sequence massive  stars which  oscillate with a few
identified modes. Identification of the detected  modes requires a high signal to noise
which is made available due to the large amplitudes of these opacity-driven
modes. On the other hand, these stars oscillate with low  frequencies 
lying near/in the dense part of the spectrum where  p modes, mixed modes and g
modes can be encountered. While this is a great advantage in order  to probe the inner
layers of the star, resolution and precise measurement of quite close  frequencies in a Fourier
spectrum requires  very  long time series. This explains the yet still 
 small number of $\beta$ Cephei stars for which a successful seismic analysis has been obtained, 
 despite the appealing prospects that a better knowledge of their structure bring up 
  valuable constrains on  their still poorly understood life end as supernovae.
  It is expected that   
 the space  experiments CoRoT (Michel {\sl et al.} 2008) and Kepler (Christensen-Dalsgaard \etal,
 2008) will  increase 
 the number of O-B  stars for which fruitful seismic analyses can be carried out as well as 
  possibly enlarge the sample to fast
 rotators.  Mode identification can be at first difficult to perform 
for fast rotators 
 but some of these fast rotating stars 
 might also  show  oscillations  of solar-like type which characteristics could help the mode
 identification. This interesting perspective has recently emerged  with 
 the discovery of the first chimera star    with the CoRoT mission (Belkacem {\sl et al.}, 2009).

\medskip
\,
\medskip
\ni{\large \bf Appendix :  Differential rotation}

\ni The expression for the  mode splitting of adiabatic nonradial
 oscillations due to a  differential rotation  $\Omega(r,\theta)$ 
can be put   into the  compact form (Hansen {\sl et al.} (1977), DG91, DG92, 
Schou {\sl et al.} (1994a,b), Pijpers (1997), CD03): 
\begin{equation}
\delta \omega_m=  m~ \int_0^R \int_0^\pi  {\cal K}_{m}(r,\theta) ~\Omega(r,\theta) ~ d\theta dr
\label{split2}
\end{equation}
where  ${\cal K}_m$ is called {\it rotational kernel}\index{rotational kernel}:
 \begin{eqnarray}
{\cal K}_{m}(r,\theta)&=& - {  ~\rho_0 r^2 \over I}
{\sin \theta \over 2}    ~\Omega(r,\theta)   \times \nonumber \\
& & \int {d\phi \over 2 \pi} ~\Bigl[ \Bigr.  \Bigl(|\xi_r|^2  -(\xi^*_r \xi_h +cc) \Bigr) ~ |Y_\ell^m|^2
   \\
&+&|\xi_h|^2 
\Bigl( \nabla_H Y_\ell^{m *}   \cdot \nabla_H Y_\ell^{m}
-{\partial |Y_\ell^m|^2 \over \partial \theta} {\cos \theta \over \sin
\theta  }  \Bigr) 
\Bigl. \Bigr]      \nonumber 
\label{split3}
\end{eqnarray}
where the spherical harmonics $ Y_\ell^m $ are normalized such that 
$$\int (Y_\ell^{m'})^*(\theta,\phi) ~  Y_\ell^m (\theta,\phi)~ {d\Omega \over 4 \pi} = \delta_{\ell,\ell'}\delta_{m,m'} $$
where $d\Omega=\sin \theta d\theta d\phi$ is the  solid angle elemental
variation   and $\delta_{\ell,\ell'} $ is the
 Kroenecker symbol.
 Mode inertia $I$ is given by 
\begin{equation}
I=  \int_0^R  \Bigl(|\xi_r|^2  + \Lambda ~|\xi_h|^2 \Bigr)  ~\rho_0 r^2  dr
\label{inertia2}
\end{equation}
with $\Lambda = \ell(\ell+1)$.

It is convenient to assume  a rotation of the type:
\begin{equation}
\Omega(r,\theta)=  \sum_{s=0}^{s_{max}} ~ \Omega_{2s}(r) ~(\cos\theta)^{2s}
\label{latit}
\end{equation}
where $\theta$ is the colatitude. 
Eq.\ref{split2} becomes:


 \begin{eqnarray}
\delta \omega_m&=& - { m\over I}\sum_{s=0}^{s_{max}} ~ \int_0^R ~ \Omega_{2s}(r) ~   ~  \times \\
& & \Bigl[ \Bigr.  ~\Bigl(|\xi_r|^2  -(\xi^*_r \xi_h +cc) \Bigr) ~ {\cal S}_s
+|\xi_h|^2  (B_1+B_2) 
\Bigl. \Bigr]   ~\rho_0 r^2  dr    \nonumber
 \label{split7}
\end{eqnarray}
where we have  defined 
\begin{equation}
{\cal S}_s \equiv  \int    ~  |Y_\ell^m|^2 ~(\cos\theta)^{2s} ~ {d\Omega \over 4 \pi}  
=\int_0^1 \mu^{2s} ~|Y_{\ell}^m(\theta,\phi)|^2   d\mu 
\label{Ss}
\end{equation}
with $\mu =\cos \theta$ and
 \begin{eqnarray}
B_1 &=& \int   ~  
\Bigl( \nabla_H Y_\ell^{m *}   \cdot \nabla_H Y_\ell^{m} \Bigr) (\cos\theta)^{2s}  ~{d\Omega \over 4 \pi}   \\
B_2&=& -\int ~( {\partial |Y_\ell^m|^2 \over \partial \theta} {\cos \theta \over \sin\theta  } 
\Bigr) ~(\cos\theta)^{2s}  ~ {d\Omega \over 4 \pi} 
\label{split8}
\end{eqnarray}
 The term in $ |\xi_h|^2$  requires a little care. Consider first $B1$. Integration by part leads to 

\begin{eqnarray}
B1 &=& - \int {d\Omega \over 4 \pi}   ~   Y_\ell^{m *} ~  \\
&~& \Bigl[ \Bigl( \nabla^2_H Y_\ell^{m} \Bigr)  ~(\cos\theta)^{2s}
+(\nabla_H Y_\ell^{m} \Bigr) ~ \cdot ~ \nabla_H((\cos\theta)^{2s}) \Bigr] \nonumber
\end{eqnarray}
Recalling that $ \nabla^2_H Y_\ell^{m} = -\Lambda ~Y_\ell^{m}  $, one gets 

\begin{eqnarray}
B1 &=&  \Lambda  ~{\cal S}_s 
- {1\over 2 }\int ~ 
 ~\Bigl[ Y_\ell^{m *}~  {\partial  Y_\ell^{m}\over \partial \theta}  
  ~  {d(\cos\theta)^{2s}) \over d\theta}  + cc \Bigr] ~ {d\Omega \over 4 \pi}
\end{eqnarray}
where cc means complexe conjugate.
Again an integration by part yields 
\begin{eqnarray}
B1 &=&   \Lambda  ~{\cal S}_s
+{1\over 2 } \int    |Y_\ell^{m}|^2 
  ~  {d \over d\theta}   \Bigl[  \sin \theta ~  {d(\cos\theta)^{2s}) \over d\theta}  
   \Bigr]~{d \theta \over 2 }
\end{eqnarray}
One finally obtains 
\begin{eqnarray}
B1 &=&  \Lambda  ~{\cal S}_s + s \Bigl[ (2s-1)  
  {\cal S}_{s-1}-     (2s+1) {\cal S}_{s} \Bigr] 
\label{B1}
\end{eqnarray}

Turning to the second term $B_2$ in Eq.\ref{split7},
an integration by part yields 
 \begin{eqnarray}
B_2 =- (2s+1) ~ {\cal S}_s     
\label{split8}
\end{eqnarray}

Inserting expressions Eq.\ref{B1} and \ref{split8} into
Eq. \ref{split7}, one obtains
 \begin{eqnarray}
  \delta \omega_m&=& m~ \sum_{s=0}^{s_{max}} ~  \int_0^R ~ \Omega_{2s}(r) ~K_{m,s}(r) dr
\label{split21}
 \end{eqnarray}
with
 \begin{eqnarray} 
 K_{m,s}(r) & = &  K(r) ~ {\cal S}_s - {1\over I}~\rho_0 r^2 ~|\xi_h|^2  s ~\Bigl[ (2s-1) 
    {\cal S}_{s-1}-     (2s+3) {\cal S}_{s} \Bigr]  
\label{split22}
 \end{eqnarray}
and \begin{equation}
 K(r)= - {1\over I}~\Bigl[|\xi_r|^2  -(\xi^*_r \xi_h +cc)+ |\xi_h|^2 
  (\Lambda-1)\Bigr]~\rho_0~ r^2
\label{split23}   \end{equation} 
Expression Eq.\ref{split22} is equivalent to Eq.25 in DG92.  
For any $s$, $ {\cal S}_s$  is given by a recurrent relation (Eq.31 in DG92). 
Note that $\delta \omega_m =\delta \omega_{-m} $. Let define the generalized splitting 
$$S_m = {\omega_m-\omega_{-m}\over 2 m } = {\delta \omega_m - \delta \omega_{-m} \over 2m} = {\delta \omega_m\over m} $$
We limit the expression for the rotation to $s_{max}=2$ i.e.:
\begin{equation}
\Omega(r,\theta)=   \Omega_{0}(r)  + \Omega_{2}(r) ~\cos^{2}\theta+ \Omega_{4}(r) ~\cos^{4}\theta
\label{latit2}
\end{equation}
then for adiabatic oscillations ($\xi_r(r) $ and $\xi_h(r)$ are real):
\begin{eqnarray}
K_{m,0}(r)  & =&  K(r) \nonumber \\
K_{m,1}(r) &=&  K(r) ~ {\cal S}_1 -{1\over I}~\xi^2_h   ~\Bigl[1- 5 {\cal S}_{1} \Bigr] ~\rho_0 r^2 \\
K_{m,2}(r) &=& K(r)~ {\cal S}_2   
-{1\over I}~\xi^2_h ~ 2 ~\Bigl[ 3   {\cal S}_{1}-  7 {\cal S}_{2} \Bigr]  ~\rho_0 r^2\nonumber
 \end{eqnarray} 
where we have used 
 ${\cal S}_{-1}=0; {\cal S}_{0}=1$.

We obtain a formulation  for the generalized splittings with a $m$ dependence of the form:
\begin{eqnarray}
& &S_m  =        \int_0^R ~   \Bigl( \Omega_{0} + \Omega_{2} 
  {\cal S}_1 + \Omega_{4} {\cal S}_2 \Bigr) ~ K(r) ~dr   \nonumber  \\
 &-&  {1  \over I}   ~ \int_0^R  \Bigl(\Omega_{2} (1-5 {\cal S}_1)  + \Omega_{4}  ~2 (3{\cal S}_{1}-7 {\cal S}_2)
   \Bigr)   ~ \xi^2_h ~\rho_0 r^2 dr 
   \label{split30}
 \end{eqnarray}



One  needs $S_1$ and $S_2$ (computed from Eq.31 in DG92): 
\begin{eqnarray}
 {\cal S}_{1} &=& {1\over 4\Lambda-3} (-2m^2+2\Lambda-1) = {2\Lambda-1\over 4\Lambda-3} - m^2~ {2\over 4\Lambda-3}  \nonumber \\
{\cal S}_2 &=& {1\over 4\Lambda-15} ~{3\over 2}~\left[{\cal S}_{1}  (-2m^2+2\Lambda-9)+1 \right]  \nonumber 
\label{SS5}
\end{eqnarray}

 \medskip
The first term in brackets in Eq.\ref{split30}  becomes 
\begin{equation} 
 \Bigl( \Omega_{0} + \Omega_{2}  {\cal S}_1 + \Omega_{4} {\cal S}_2 \Bigr)  =\Omega_{0} 
 ~(1+  R_0+m^2 ~R_1 +m^4~R_2) 
 \label{RR}
\end{equation}
where 
\begin{eqnarray} 
R_0 &=& {\Omega_{2} \over \Omega_0} ~ { 2\Lambda-1 \over 4\Lambda-3} + 3~{ \Omega_{4}\over \Omega_0} 
 ~{\left[  ( 2\Lambda^2 -8 \Lambda +3) \right]\over (4\Lambda-15)(4\Lambda-3)}     \nonumber \\
R_1  &=& -  { 2 \over 4\Lambda-3} ~\Bigl[  {\Omega_{2}  \over \Omega_0} ~  + 3~ {\Omega_4\over \Omega_0} ~  
{(2\Lambda-5)\over (4\Lambda-15)} \Bigr] \\
R_2 &=&   {\Omega_4\over \Omega_0} ~  {6\over (4 \Lambda-15)(4\Lambda-3)}   \nonumber 
   \label{SS81}
 \end{eqnarray}

For the second term in Eq.\ref{split30}, one has:
 \begin{equation} 
 \Omega_{2} (1-5 {\cal S}_1)  + \Omega_{4}  ~2 (3{\cal S}_{1}-7 {\cal S}_2  = \Omega_0~(Q_0 +m^2 Q_2+m^4 Q_2) 
 \label{QQ} 
 \end{equation}
where
\begin{eqnarray} 
Q_0 &=&{2\over 4\Lambda-3} ~\Bigl[ {\Omega_2 \over \Omega_0} (1 - 3 \Lambda) 
- 6  {\Omega_4\over \Omega_0} {( 3\Lambda^2 -11 \Lambda +3) \over 4\Lambda-15} \Bigr]
\nonumber \\
Q_1 &=& ~ {10\over 4\Lambda-3}   ~\Bigl[ {\Omega_{2}\over \Omega_0}+ 12 {\Omega_{4}\over \Omega_0} {(\Lambda-2)\over (4 \Lambda-15)}\Bigr]
   \\
Q_2 =& -&  {4\over (4\Lambda-3)} ~  {21\over (4\Lambda-15)} ~ {\Omega_{4}\over \Omega_0}\nonumber     
\label{SS82}
 \end{eqnarray}

\medskip
Collecting terms from Eq.\ref{RR} and Eq.\ref{QQ}, the generalized splitting Eq.\ref{split30} takes the expression:





 \begin{eqnarray}
S_m  &=&  \int_0^R   \Omega_0(r) ~K(r) ~dr+ \sum_{s=0}^{s=2}  m^{2s} ~ H_{s}(\Omega) 
\label{sm84}
 \end{eqnarray}
with 
  \begin{eqnarray}
H_{s} (\Omega) =    \int_0^R  \Omega_0(r)~ \Bigl[ R_s ~K(r) -Q_s ~ {1 \over I}\xi^2_h   \Bigr] ~ \rho_0 r^2
dr 
 \end{eqnarray}

\medskip

For a depth independent rotation law, $\Omega(\theta)$,  
 $\Omega_{2j}, j=0,2$ are depth independent and  $R_s$ and $Q_s$ are constant. 
then for  a triplet $\ell=1$ ($\Lambda=2$):  
 \begin{eqnarray}
S_1  &=&   \Omega_0 ~ \beta  + \Omega_0~ (\sum_{s=0}^{s=2}   ~ R_{s}(\Omega)) ~\beta + ( \sum_{s=0}^{s=2} 
Q_{s}(\Omega)) ~ \gamma ~~~~~
 \end{eqnarray}
with

\begin{eqnarray}  
 \sum_{s=0}^{s=2} R_s &=& {1\over 5} ~  {\Omega_{2} \over \Omega_{0}} +  {3\over 7} ~  {\Omega_{4} \over
\Omega_{0}} \\
\sum_{s=0}^{s=2} Q_s &=&   - {24\over 5}~  {\Omega_{4}\over\Omega_{0}}   
  \end{eqnarray} 
and
\begin{eqnarray} 
\beta   &=& \int_0^R ~K(r) ~dr\\
\gamma &=& -{1 \over I}~ \int_0^R  ~\xi^2_h ~ \rho_0 r^2~dr
\label{gama} 
\end{eqnarray}

\end{document}